\documentclass{bmcart}

\usepackage{amsthm,amsmath}
\usepackage[utf8]{inputenc} 

\usepackage{color}

\usepackage[caption=false]{subfig}

\startlocaldefs
\usepackage{graphicx}
\usepackage{amssymb}
\usepackage{amsxtra}
\usepackage{bm}
\usepackage{algorithm}
\usepackage{algorithmic}
\makeatletter
 \def\Hline{%
 \noalign{\ifnum0=`}\fi\hrule \@height 1.4pt \futurelet
 \reserved@a\@xhline}
\makeatother
\usepackage{cite,url}
\DeclareMathOperator{\STFT}{STFT}
\DeclareMathOperator{\ISTFT}{ISTFT}
\endlocaldefs

\begin{document}

\begin{frontmatter}

\begin{fmbox}
\dochead{Research}


\title{Consistent independent low-rank matrix analysis for determined blind source separation}


\author[
   addressref={aff1},                   
   corref={aff1},                       
   email={d-kitamura@ieee.org}   
]{\inits{DK}\fnm{Daichi} \snm{Kitamura}}
\author[
   addressref={aff2},
   noteref={n1},
   email={k.yatabe@asagi.waseda.jp}
]{\inits{KY}\fnm{Kohei} \snm{Yatabe}}


\address[id=aff1]{
  \orgname{National Institute of Technology, Kagawa College}, 
  \street{355 Chokushi},                     %
  \postcode{761-8058}                                
  \city{Takamatsu, Kagawa},                              
  \cny{Japan}                                    
}
\address[id=aff2]{%
  \orgname{Waseda University},
  \street{3-4-1 Okubo},
  \postcode{169-8555}
  \city{Shinjuku-ku, Tokyo},
  \cny{Japan}
}


\begin{artnotes}
\note[id=n1]{Equal contributor} 
\end{artnotes}

\end{fmbox}


\begin{abstractbox}

\begin{abstract}
Independent low-rank matrix analysis (ILRMA) is the state-of-the-art algorithm for blind source separation (BSS) in the determined situation (the number of microphones is greater than or equal to that of source signals).
ILRMA achieves a great separation performance by modeling the power spectrograms of the source signals via the nonnegative matrix factorization (NMF).
Such a highly developed source model can solve the permutation problem of the frequency-domain BSS {to a large extent}, which is the reason for the excellence of ILRMA.
In this paper, we further improve the separation performance of ILRMA by additionally considering the general structure of spectrograms, which is called \textit{consistency}, and hence we call the proposed method \textit{Consistent ILRMA}.
Since a spectrogram is calculated by an overlapping window (and a window function induces spectral smearing called main- and side-lobes), the time-frequency bins depend on each other.
In other words, the time-frequency components are related to each other via the uncertainty principle.
Such co-occurrence among the spectral components can function as an assistant for solving the permutation problem, which has been demonstrated by a recent study.
On the basis of these facts, we propose an algorithm for realizing Consistent ILRMA by slightly modifying the original algorithm.
Its performance was extensively evaluated through experiments performed with various window lengths and shift lengths.
The results indicated several tendencies of the original and proposed ILRMA that include some topics not fully discussed in the literature.
For example, the proposed Consistent ILRMA tends to outperform the original ILRMA when the window length is sufficiently long compared to the reverberation time of the mixing system.
\end{abstract}




\end{abstractbox}
%

\end{frontmatter}




\vspace{2pt}
\section{Introduction}
\vspace{1pt}

Blind source separation (BSS) is a technique for separating individual sources from an observed mixture without knowing how they were mixed.
BSS for multichannel audio signals observed by multiple microphones has been particularly studied~\cite{Comon1995_ica,Smaragdis1998_fdica,Kurita2000_permutation,Murata2001_corPermSolver,Saruwatari2006_doaPermSolver,Sawada2004_corDoaPermSolver,Hiroe2006_iva,TKim2006_iva,Kim2007_iva,Ono2011_auxiva,Kitamura2016_ilrma,Kitamura2018_ilrma,Tachikawa2018}.
The BSS problem can be divided into two situations: underdetermined (the number of microphones is less than the number of sources) and (over-)determined (the number of microphones is greater than or equal to the number of sources) cases.
This paper focuses on the determined BSS problem, as high-quality separation can be achieved compared with the underdetermined BSS methods.

Independent component analysis (ICA) is the most popular and successful algorithm for solving the determined BSS problem~\cite{Comon1995_ica}.
It estimates a demixing matrix (the inverse system of the mixing process) by assuming statistical independence between the sources.
For a mixture of audio signals, ICA is usually applied in the time-frequency domain via the short-time Fourier transform (STFT) because the sources are mixed up by convolution.
This strategy is called frequency-domain ICA (FDICA)~\cite{Smaragdis1998_fdica} and independently applies ICA to the complex-valued signals in each frequency.
Then, the estimated frequency-wise demixing matrices must be aligned over all frequencies so that the frequency components of the same source are grouped together.
Such alignment of the frequency components is called a \textit{permutation problem}~\cite{Kurita2000_permutation,Murata2001_corPermSolver,Saruwatari2006_doaPermSolver,Sawada2004_corDoaPermSolver} and a complete solution to it has not been established.
Therefore, a great deal of research has tackled this problem.

{To avoid the permutation misalignment as much as possible}, various sophisticated source models have been proposed.
Independent vector analysis (IVA)~\cite{Hiroe2006_iva,TKim2006_iva,Kim2007_iva,Ono2011_auxiva} is one of the most successful methods in the early stage of the development.
It assumes higher-order dependences (co-occurrence among the frequency components) of each source by utilizing a spherical generative model of the source frequency vector.
This assumption enables IVA to simultaneously estimate the frequency-wise demixing matrices and solve the permutation problem {to a large extent} using only one objective function.
It has been further developed by improving its source model.
One natural and powerful extension of IVA is independent low-rank matrix analysis (ILRMA)~\cite{Kitamura2016_ilrma,Kitamura2018_ilrma}, which integrates the source model of nonnegative matrix factorization (NMF)~\cite{Lee1999_Nature,Lee2000} based on the Itakura--Saito divergence (IS-NMF)~\cite{Fevotte2009} into IVA.
This extension has greatly improved the performance of separation by taking the low-rank time-frequency structure (co-occurrence among the time-frequency bins) of the source signals into account.
ILRMA has achieved the state-of-the-art performance and been further developed by several researchers~\cite{Mitsui2017_icassp,Kagami2018_icassp,Ikeshita2018_icassp,Kitamura2018_superGaussIlrma,Yoshii2018_ilrta,Ikeshita2018_ipsdta,Ikeshita2019,Makishima2019_idlma,Sekigushi2019,Mogami2019_subGaussIlrma,Takahashi2020_icassp,Togami2020_icassp,Kanoga2020_neuro}.
In this respect, ILRMA can be considered the new standard of the determined BSS algorithms.
{However, the separation performance of IVA and ILRMA is still inferior compared to the ideal performance of ICA-based frequency-domain BSS. 
In~\cite{Kitamura2017_ilrma}, the performances of IVA and ILRMA were compared with that of FDICA with perfect permutation alignment using reference sources (ideal permutation solver), and it was confirmed that there is still a noticeable room for improvement of ILRMA-based BSS.
In fact, IVA and ILRMA often encounter the block permutation problem, that is, group-wise permutation misalignment of components between sources~\cite{Liang2012_blockPermIVA}.}

The \textit{consistency} of a spectrogram is another promising approach for solving the permutation problem.
A recent study has shown that STFT can provide some effective information related to the co-occurrence among the time-frequency bins~\cite{Yatabe2020_consistIca}.
Since an overlapping window is utilized in STFT, the time-frequency bins are related to each other based on the overlapping segments.
The frequency components within a segment are also related to each other because of the spectral smearing called main- and side-lobes of the window.
In other words, the time-frequency components are not independent but related to each other via the \textit{uncertainty principle} of time-frequency representation.
Such relations have been well-studied in phase-aware signal processing~\cite{Gerkmann15,Mowlaee16,MowlaeeBook,YatabePC,YatabeReLU,MasuyamaManifold,MasuyamaADMM,MasuyamaDeGLI,MasuyamaHPSS,MasuyamaLowRank,Masuyama2020} by the name of spectrogram consistency~\cite{LeRouxGLA,LeRouxWiener,fastGLA,YatabeAST2019}.
In the previous study~\cite{Yatabe2020_consistIca}, the spectrogram consistency was imposed on BSS to help the algorithm solve the permutation problem.
This is an approach very different from the conventional studies of determined BSS because it utilizes the general property of STFT \textit{independent of the source model} (in contrast to the above-mentioned methods that focused on modeling of the source signals without considering the property of STFT).
As the spectrogram consistency can be incorporated with any source model, its combination with the state-of-the-art algorithm should achieve a high separation performance.

However, the paper that proposed the combination of consistency and determined BSS~\cite{Yatabe2020_consistIca} only showed the potential of consistency in an experiment using FDICA and IVA.
The paper claimed that it was a first step of incorporating the spectrogram consistency with determined BSS, and no advanced method was tested.
In particular, ILRMA was not considered because its algorithm is far more complicated than that derived in~\cite{Yatabe2020_consistIca}, and thus it is not clear whether (and how much) the spectrogram consistency might improve the state-of-the-art BSS algorithm.

In this paper, we propose a new variant of ILRMA called \textit{Consistent ILRMA} that considers the spectrogram consistency within the algorithm of ILRMA.
The combination of IS-NMF and spectral smoothing of the inverse STFT (see {Figs.\:\ref{fig:consistSpectrogram} and \ref{fig:consistSpectrogramMusSpe} in Sect.~\ref{sec:convSolv}}) achieves the source modeling for a complex spectrogram.
In particular, the spectral smearing in the frequency direction ties the adjacent frequency bins together, and this effect of spectrogram consistency helps ILRMA to solve the permutation problem.
Since consistency is a concept depending on the parameters related to a window function, we extensively tested the separation performance of Consistent ILRMA through experiments with various window lengths and shift lengths.
The results clarified several tendencies of the conventional and proposed methods, including that the proposed method outperforms the original ILRMA when the window length is sufficiently long compared to the reverberation time of the mixing system.

\section{Permutation problem of frequency-domain BSS and spectrogram consistency}
\subsection{Formulation of frequency-domain BSS}

Let the $l$th sample of a time-domain signal be denoted as $x[l]$, and $N$ source signals be observed by $M$ microphones.
Then, the $l$th sample of the multichannel source, observed, and separated signals are respectively denoted as
\begin{align}
    \bm{s}[l] &= \bigl[\, s_1[l], s_2[l], \cdots, s_n[l], \cdots s_N[l] \,\bigr]^\mathrm{T}
    \in \mathbb{R}^{N}, \label{eq:sVecTime} \\
    \bm{x}[l] &= \bigl[\, x_1[l], x_2[l], \cdots, x_m[l], \cdots x_M[l] \,\bigr]^\mathrm{T}
    \in \mathbb{R}^{M}, \label{eq:xVecTime} \\
    \bm{y}[l] &= \bigl[\, y_1[l], y_2[l], \cdots, y_n[l], \cdots y_N[l] \,\bigr]^\mathrm{T}
    \in \mathbb{R}^{N}, \label{eq:yVecTime}
\end{align}
where $n=1, \cdots, N$, $m=1, \cdots, M$, and $l=1, \cdots, L$ are the indexes of sources, microphones (channels), and discrete time, respectively, and $\cdot^\mathrm{T}$ denotes the transpose.
BSS aims at recovering the source signal $\bm{s}$ from the observed signal $\bm{x}$, i.e., making $\bm{y}$ as close to $\bm{s}$ as possible.

In the frequency-domain BSS, those signals are handled in the time-frequency domain via STFT.
Let the window length and shifting step of STFT be denoted as $Q$ and $\tau$, respectively.
Then, the $j$th segment of a signal $z[l]$ is defined as
\begin{align}
    \bm{z}^{[j]} &= \bigl[\, z[(j\!-\!1)\tau \!+\!1], z[(j\!-\!1)\tau \!+\!2], \cdots, z[(j\!-\!1)\tau \!+\!Q] \,\bigr]^\mathrm{T}\!, \nonumber \\
    &= \bigl[\, z^{[j]}[1], z^{[j]}[2], \cdots, z^{[j]}[q], \cdots, z^{[j]}[Q] \,\bigr]^\mathrm{T} \in \mathbb{R}^{Q}, \label{eq:shortTimeVec}
\end{align}
where $j=1,\cdots,J$ and $q=1,\cdots,Q$ are the indexes of the segments and in-segment samples, respectively, and the number of segments is given by $J=L/\tau$ with some zero-padding for adjusting the signal length $L$ if necessary.
STFT of a signal $\bm{z} = [\,z[1],\cdots,z[L]\,]^\mathrm{T}\in\mathbb{R}^L$ is denoted by
\begin{equation}
    \bm{Z} = \STFT_{\bm{\omega}}(\bm{z})
    \;\;\in\mathbb{C}^{I\times J},
\end{equation}
where the $(i,j)$th bin of the spectrogram $\bm{Z}$ is given as
\begin{equation}
    z_{ij} = \sum_{q=1}^Q \omega[q]\,z^{[j]}[q]\;\mathrm{e}^{-\imath2\pi(q-1)(i-1)/F},
\end{equation}
$i=1,\cdots,I$ is the index of frequency bins, $F$ is an integer satisfying $\lfloor F/2 \rfloor+1=I$, $\lfloor\cdot\rfloor$ is the floor function, $\imath$ denotes the imaginary unit, and $\bm{\omega}$ is an analysis window. 
The inverse STFT with a synthesis window $\widetilde{\bm{\omega}}$ is also defined in the usual way and denoted as $\ISTFT_{\widetilde{\bm{\omega}}}(\cdot)$.
In this paper, we assume that the window pair satisfies the following perfect reconstruction condition:
\begin{equation}
    \bm{z} = \ISTFT_{\widetilde{\bm{\omega}}}(\STFT_{\bm{\omega}}(\bm{z}))\qquad\forall \bm{z}\in\mathbb{R}^L.
    \label{eq:perfectReconstructionCond}
\end{equation}

By applying STFT, the $(i,j)$th bin of the spectrograms of source, observed, and separated signals can be written as
\begin{align}
    \bm{s}_{ij} &= \left[\, s_{ij1}, s_{ij2}, \cdots, s_{ijn}, \cdots s_{ijN} \,\right]^\mathrm{T}
    \in \mathbb{C}^{N}, \label{eq:sVec} \\
    \bm{x}_{ij} &= \left[\, x_{ij1}, x_{ij2}, \cdots, x_{ijm}, \cdots x_{ijM} \,\right]^\mathrm{T}
    \in \mathbb{C}^{M}, \label{eq:xVec} \\
    \bm{y}_{ij} &= \left[\, y_{ij1}, y_{ij2}, \cdots, y_{ijn}, \cdots y_{ijN} \,\right]^\mathrm{T}
    \in \mathbb{C}^{N}. \label{eq:yVec}
\end{align}
We also denote the spectrograms corresponding to the $n$th or $m$th signals in \eqref{eq:sVec}--\eqref{eq:yVec} as $\bm{S}_n\in\mathbb{C}^{I\times J}$, $\bm{X}_m\in\mathbb{C}^{I\times J}$, and $\bm{Y}_n\in\mathbb{C}^{I\times J}$, whose elements are $s_{ijn}$, $x_{ijm}$, and $y_{ijn}$, respectively.
In the ordinary frequency-domain BSS, an instantaneous mixing process for each frequency bin is assumed:
\begin{align}
    \bm{x}_{ij} = \bm{A}_i\bm{s}_{ij}, \label{eq:mixingSystem}
\end{align}
where $\bm{A}_i\in\mathbb{C}^{M\times N}$ is a frequency-wise mixing matrix. 
The mixture model \eqref{eq:mixingSystem} is approximately valid when the reverberation time is sufficiently shorter than the length of the analysis window used in STFT~{\cite{Kowalski2010}}.

Hereafter, we consider the determined case, i.e., $M=N$.
In this case, BSS can be achieved by estimating the inverse of $\bm{A}_i$ for all frequency bins.
By denoting an approximate inverse as $\bm{W}_i\approx\bm{A}_i^{-1}$, the separation process can be written as
\begin{align}
    \bm{y}_{ij} = \bm{W}_i\bm{x}_{ij}, \label{eq:demixingSystem}
\end{align}
where $\bm{W}_i=[\bm{w}_{i1},\bm{w}_{i2},\cdots,\bm{w}_{iN}]^\mathrm{H}\in\mathbb{C}^{N\times M}$ is a frequency-wise demixing matrix and $\cdot^\mathrm{H}$ denotes the Hermitian transpose.
The aim of a determined BSS algorithm is to find the demixing matrices for all frequency bins so that the separated signals approximate the source signals.

\subsection{Permutation problem in determined BSS}

In practice, the scale and  permutation of the separated signals are unknown because the information of the mixing process is missing.
That is, when the separation is correctly performed by some demixing matrix $\bm{W}_i$ as in \eqref{eq:demixingSystem}, the following signal is also a solution to the BSS problem:
\begin{equation}
    \hat{\bm{y}}_{ij} = \hat{\bm{W}}_i\bm{x}_{ij}
    \qquad
    (\hat{\bm{W}}_i = \bm{D}_i\bm{P}_i\bm{W}_i), \label{eq:indeterminancy}
\end{equation}
where $\bm{D}_i\in\mathbb{C}^{N\times N}$ and $\bm{P}_i\in\{0,1\}^{N\times N}$ are arbitrary diagonal and permutation matrices, respectively.
While the signal scale can easily be recovered by applying the back projection~\cite{Matsuoka2001_PB}, the permutation of the estimated signals $\hat{\bm{y}}_{ij}$ must be aligned for all frequency bins, i.e., $\bm{P}_i$ must be the same for all $i$.
This alignment of the permutation of estimated signals is the permutation problem, which is the main obstacle of the frequency-domain determined BSS.

In FDICA, a permutation solver (realignment process of $\bm{P}_i$) is utilized as a post-processing applied to the frequency-wise separated signals $\hat{\bm{y}}_{ij}$~\cite{Murata2001_corPermSolver,Saruwatari2006_doaPermSolver,Sawada2004_corDoaPermSolver}.
In recent frequency-domain BSS methods, an additional assumption on sources (or source model) is introduced to circumvent the permutation problem.
For example, IVA assumes simultaneous co-occurrence of all frequency components in the same source, and ILRMA assumes a low-rank structure of the power spectrogram $\bm{Y}_n$.
Other source models have also been proposed for improving the separation performance~\cite{Yatabe2018_pdsBss,Yatabe2019_tfmBss,Yatabe2020_tfmBss}.
These source models can avoid the permutation problem to some extent during the estimation of $\hat{\bm{W}}_i$.
Recent developments of determined BSS have been achieved via the quest to find a better source model that represents the source signals more precisely.

\subsection{Solving permutation problem by spectrogram consistency} \label{sec:convSolv}
A recent paper reported another approach for solving the permutation problem based on the general property of STFT called spectrogram consistency~\cite{Yatabe2020_consistIca}.
The consistency is a fundamental property of a spectrogram.
Since any time-frequency representation has a theoretical limitation called the uncertainty principle, the time-frequency bins of a spectrogram are not independent but related to each other.
The inverse STFT always modifies the spectrogram $\bm{Z}_n$ that violates this kind of inter-time-frequency relation so that the relation is recovered.
That is, a spectrogram $\bm{Z}_n$ properly retains the inter-time-frequency relation if and only if
\begin{equation}
    \mathcal{E}(\bm{Z}_n) = \bm{Z}_n - \STFT_{\bm{\omega}}(\ISTFT_{\widetilde{\bm{\omega}}}(\bm{Z}_n)) \label{eq:consiste}
\end{equation}
is zero, i.e., $\|\mathcal{E}(\bm{Z}_n)\|=0$ for a norm $\|\cdot\|$.
Such spectrogram $\bm{Z}_n$ satisfying $\|\mathcal{E}(\bm{Z}_n)\|=0$ is said to be \textit{consistent}.

\begin{figure}[t]
    \begin{center}
        \includegraphics[width=0.5\columnwidth]{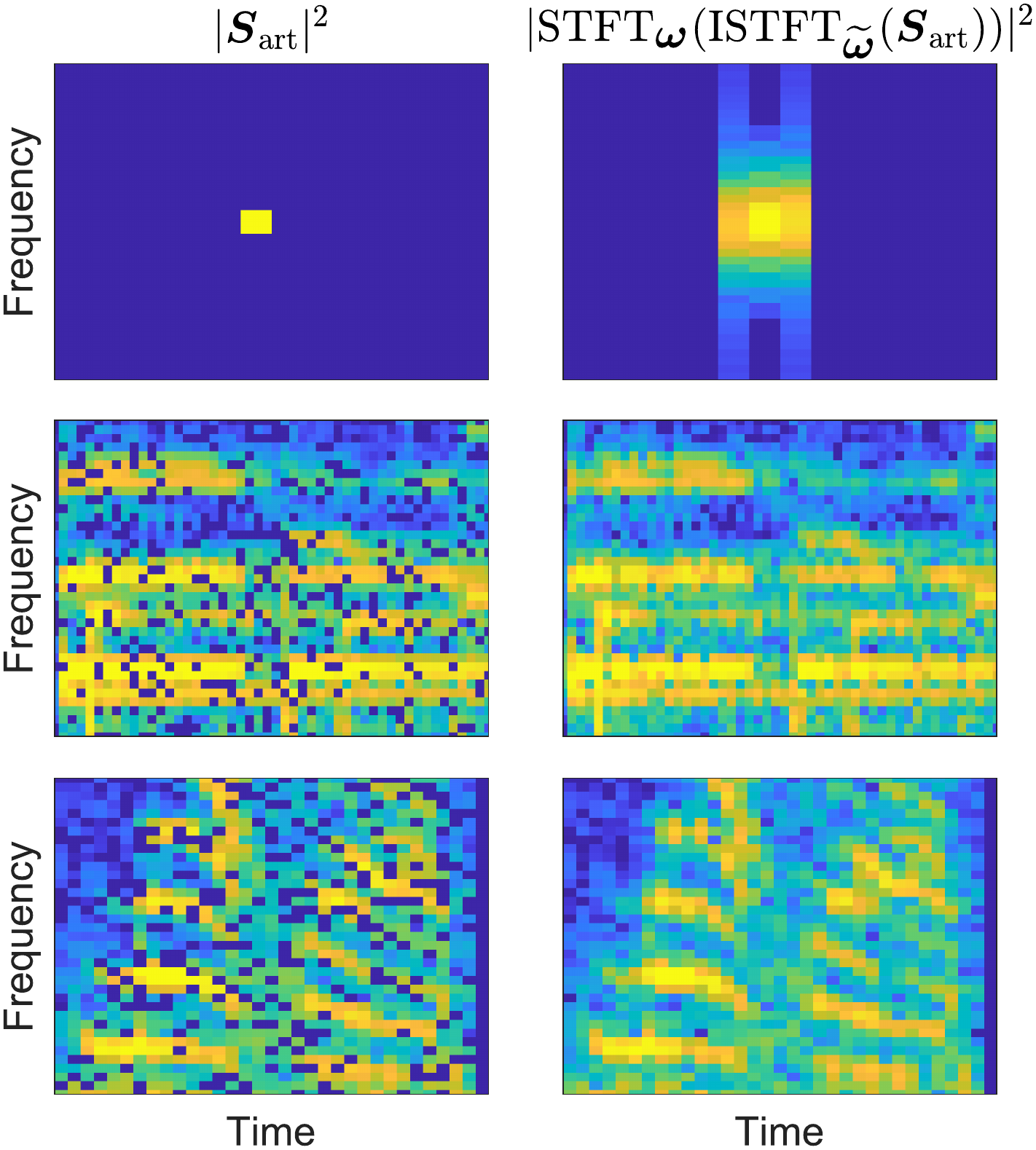}
    \end{center}
    \caption{{Inconsistent power spectrograms $|\bm{S}_\mathrm{art}|^2$ (left column) and their consistent version (right column) obtained by applying inverse STFT and STFT. The top-left spectrogram is artificially produced with random phase. The middle-left and the bottom-left spectrograms are music and speech signals with random dropout. Enforcing spectrogram consistency can be viewed as a smoothing process of the inconsistent spectrogram along both time and frequency axes.}}
    \label{fig:consistSpectrogram}
\end{figure}

{Figure\:\ref{fig:consistSpectrogram} demonstrates the effect of spectrogram consistency, where $\bm{S}_\mathrm{art}\in\mathbb{C}^{I\times J}$ is an artificially produced complex-valued spectrogram and $|\bm{S}_\mathrm{art}|^2$ is its power spectrogram. 
The notation $|\cdot|^{2}$ for a matrix input represents the element-wise squared absolute value. 
By applying $\STFT_{\bm{\omega}}(\ISTFT_{\widetilde{\bm{\omega}}}(\cdot))$, the inconsistent spectrogram $\bm{S}_\mathrm{art}$ shown in the left column of Fig.\:\ref{fig:consistSpectrogram} is converted into the corresponding consistent spectrogram, which is a smoothed version of $\bm{S}_\mathrm{art}$, as shown in the right column.
This smoothing process occurs because the main- and side-lobes of the window function (and the overlap-add process) spread the energy of a time-frequency bin.}

\begin{figure}[t]
    \centering
    \subfloat[Music signals: guitar (top row) and vocals (bottom row)]{
    \includegraphics[width=0.93\columnwidth]{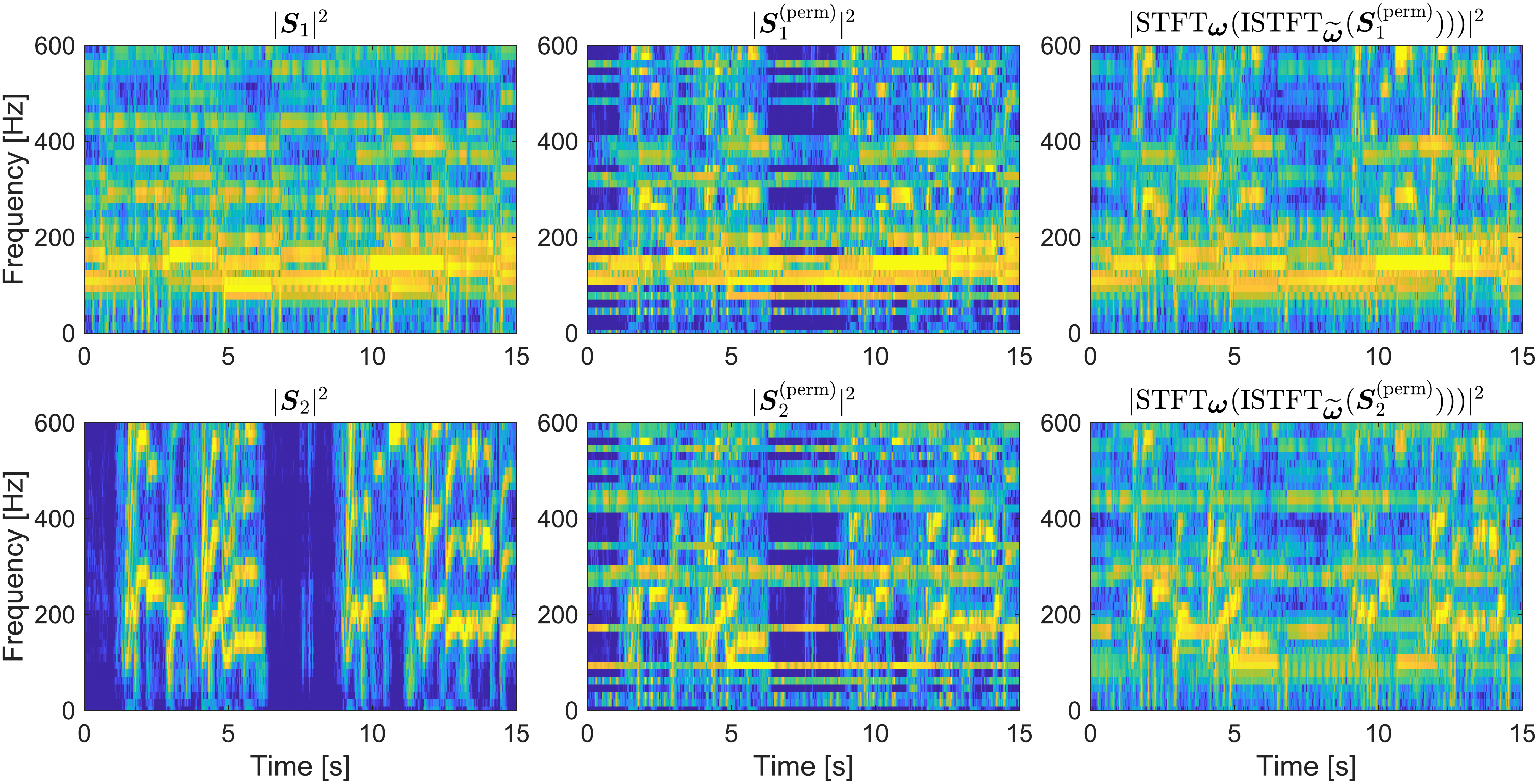}} \\
    \subfloat[Speech signals: female (top row) and male (bottom row)]{
    \includegraphics[width=0.93\columnwidth]{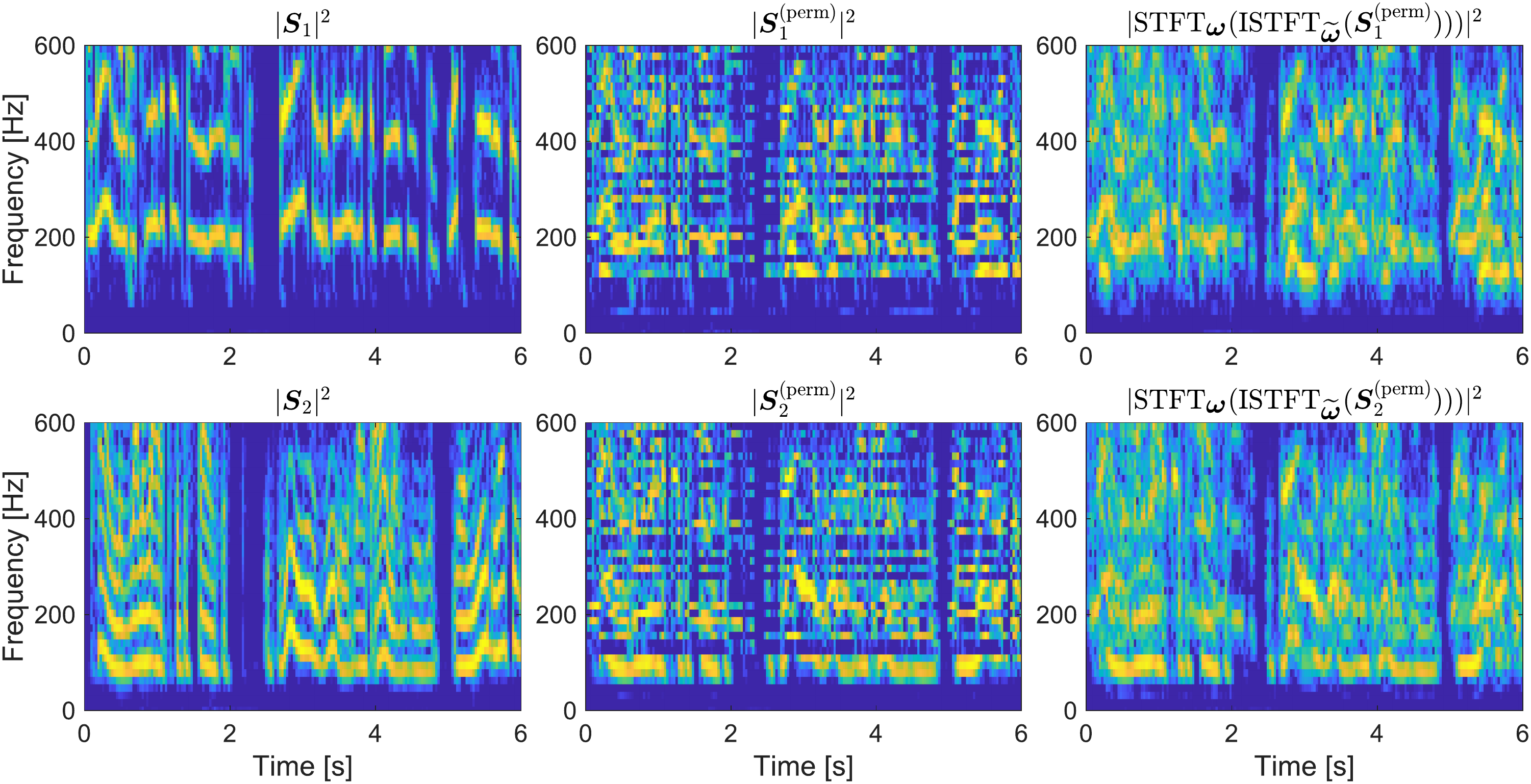}}
    \caption{{Smoothing effect of spectrogram consistency applied to permutation misaligned signals: (a) music and (b) speech. The left column shows the original source signals $|\bm{S}_n|^2$ and the center column shows their randomly permuted versions, which simulates the permutation problem and is denoted as $\bm{S}_n^{(\mathrm{perm})}$.
    The right column shows the consistent versions of $\bm{S}_n^{\mathrm{(perm)}}$. The smoothing effect mixes up the signals.}}
    \label{fig:consistSpectrogramMusSpe}
\end{figure}

{Since the inverse STFT is a process of recovering the consistency (the inter-time-frequency relation), it has the capability of aligning the frequency components.
This is also demonstrated in Fig.\:\ref{fig:consistSpectrogramMusSpe}. 
As a simulation of the permutation problem, the frequency bins in $\bm{S}_1$ and $\bm{S}_2$ were randomly shuffled to obtain the spectrogram with permutation misalignment, $\bm{S}_n^{\mathrm{(perm)}}$ (the center column in the figure), which is a typical output signal of FDICA.
Note that these misaligned spectrograms are perfectly separated for each frequency because each time-frequency bin contains only one of the two sources.
By enforcing spectrogram consistency, the smoothing process spread the time-frequency components as shown in the right column of Fig.\:\ref{fig:consistSpectrogramMusSpe}.
In other words, the inverse STFT mixes up the separated signals if the frequency-wise permutation is not aligned correctly.
Therefore, enforcing consistency within a BSS algorithm by applying $\STFT_{\bm{\omega}}(\ISTFT_{\widetilde{\bm{\omega}}}(\cdot))$ can improve the separation performance to some extent \cite{Yatabe2020_consistIca}.}

\section{Proposed method}

By incorporating spectrogram consistency into ILRMA, we propose a novel BSS method named \textit{Consistent ILRMA}.
In this section, after stating our motivation and contributions, we first review the standard ILRMA introduced in~\cite{Kitamura2016_ilrma,Kitamura2018_ilrma} and then propose the consistent version of ILRMA with an algorithm that achieves Consistent ILRMA and is openly available on the web.

{\subsection{Motivations and contributions}}
{The previous paper~\cite{Yatabe2020_consistIca} only reported that the performances of traditional BSS algorithms, FDICA and IVA, were improved by enforcing consistency during the estimation of the demixing matrix $\bm{W}_i$.
In addition, no detailed experimental analysis related to STFT parameters was provided, even though the parameters of window functions in the STFT and inverse STFT directly affect the smoothing effect of spectrogram consistency.

The spectrogram consistency is a general property of STFT, and therefore it can be combined with any source model for determined BSS.
Its combination with state-of-the-art models, including ILRMA, is of great interest because the current mainstream algorithm for determined audio source separation is centered on ILRMA, which is based on an NMF-based richer time-frequency source model.
Indeed, many recent papers are based on the framework of ILRMA~\cite{Mitsui2017_icassp,Kagami2018_icassp,Ikeshita2018_icassp,Kitamura2018_superGaussIlrma,Yoshii2018_ilrta,Ikeshita2018_ipsdta,Ikeshita2019,Makishima2019_idlma,Sekigushi2019,Mogami2019_subGaussIlrma,Takahashi2020_icassp,Togami2020_icassp,Kanoga2020_neuro}.
Even though combining ILRMA with the spectrogram consistency should be able to exceed the limit of existing BSS algorithms, no such method has been investigated in the literature.

In this paper, we propose a new BSS algorithm that combines ILRMA and spectrogram consistency.
Our first contribution is an algorithm that achieves Consistent ILRMA by inserting $\STFT_{\bm{\omega}}(\ISTFT_{\widetilde{\bm{\omega}}}(\cdot))$ into the iterative optimization algorithm of ILRMA.
The second contribution is to apply a scale-aligning process called \textit{iterative back projection} within the iterative algorithm.
This process enhances the separation performance when it is combined with spectrogram consistency.
The third contribution is an experimental finding that spectrogram consistency can work properly with the iterative back projection.
We found that both Consistent IVA and Consistent ILRMA require iterative back projection to achieve a good performance.
Our fourth contribution is to provide the massive experimental results for several window functions, window lengths, shift lengths, reverberation times, and source types.
We also provide discussions for clarifying the tendency of ILRMA with spectrogram consistency.}

\subsection{Standard ILRMA~\cite{Kitamura2018_ilrma}}

The original ILRMA~\cite{Kitamura2018_ilrma} was derived from the following generative model of the spectrograms of the separated signals:
\begin{equation}
    \bm{Y}_n \sim p(\bm{Y}_n) = \prod_{i,j} \mathcal{N}_\mathrm{c}(0,r_{ijn})
    = \prod_{i,j} \frac{ 1 }{ \pi r_{ijn} } \exp{\left( -\frac{ |y_{ijn}|^2 }{ r_{ijn} } \right)}, \label{eq:srcGenModel}
\end{equation}
where $\mathcal{N}_\mathrm{c}(\mu, r)$ is the circularly symmetric complex Gaussian distribution with mean $\mu$ and variance $r$. 
In this model, {the source component $y_{ijn}$ is assumed to obey a zero-mean and isotropic distribution, i.e., the phase of $y_{ijn}$ is generated from the uniform distribution in the range $[0,2\pi)$ and the real and imaginary parts of $y_{ijn}$ are mutually independent.
The validity of this assumption is shown in the Appendix.}
The variance $r_{ijn}$ can be viewed as an expectation value of $|y_{ijn}|^{2}$.
This variance $r_{ijn}$ as a two-dimensional array indexed by $(i,j)$ is denoted as $\bm{R}_n\in\mathbb{R}_{> 0}^{I\times J}$, which is called the variance spectrogram corresponding to the $n$th source.
In ILRMA, the variance matrix $\bm{R}_n$ is modeled using the rank-$K$ NMF, as
\begin{align}
    \bm{R}_n = \bm{T}_n\bm{V}_n, \label{eq:srcModelIlrma}
\end{align}
where $\bm{T}_n\in\mathbb{R}_{> 0}^{I\times K}$ and $\bm{V}_n\in\mathbb{R}_{> 0}^{K\times J}$ are the basis and activation matrices in NMF.
The basis vectors in $\bm{T}_n$, which represent spectral patterns of the $n$th source signal, are indexed by $k=1, \cdots, K$.
As in FDICA, statistical independence between the source signals is also assumed in ILRMA: 
\begin{equation}
    p(\bm{Y}_1, \bm{Y}_2, \cdots, \bm{Y}_N) = \prod_{n} p(\bm{Y}_n). \label{eq:independence}
\end{equation}

\begin{figure}[t]
    \begin{center}
        \includegraphics[width=0.9\columnwidth]{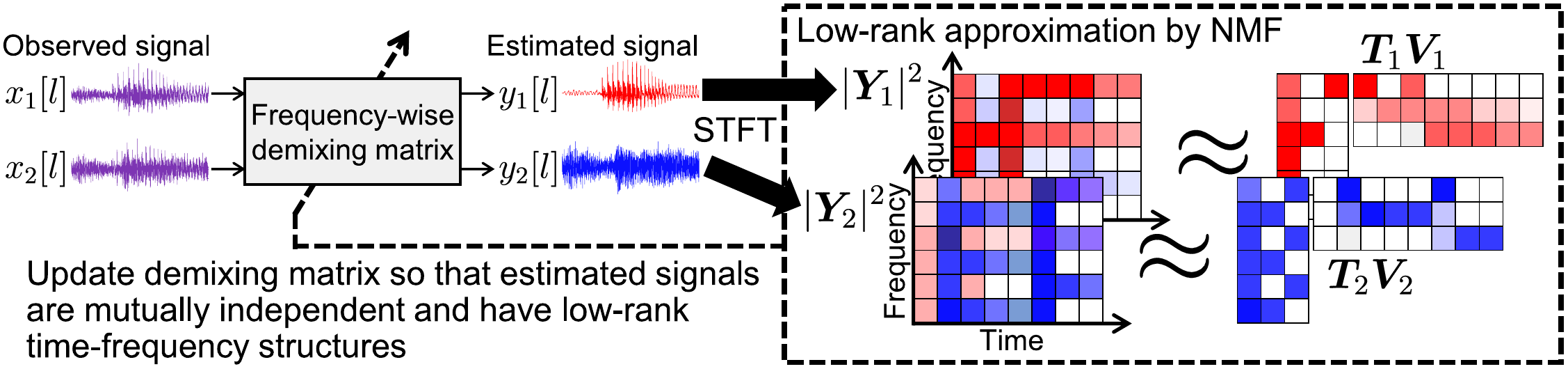}
    \end{center}
    \caption{{BSS principle of standard ILRMA.}}
    \label{fig:ILRMAfigure}
\end{figure}

ILRMA estimates the demixing matrix $\bm{W}_i$ so that the power spectrograms of the separated signals $|\bm{Y}_n|^{2}$ have a low-rank structure that can be well-approximated by $\bm{T}_n\bm{V}_n$ with small $K$.
{This BSS principle of ILRMA is illustrated in Fig.\:\ref{fig:ILRMAfigure}.}
When the low-rank source model can appropriately fit to the power spectrograms of the original source signals $|\bm{S}_n|^2$, ILRMA provides an excellent separation performance without explicitly solving the permutation problem afterward.

The demixing matrix $\bm{W}_i$ and the nonnegative matrices $\bm{T}_n$ and $\bm{V}_n$ can be obtained through maximum likelihood estimation. 
The negative log-likelihood to be minimized, denoted by $\mathcal{L}$, is given as~\cite{Kitamura2018_ilrma}: 
\begin{align}
    \mathcal{L} &= - \log p(\bm{X}_1, \bm{X}_2, \cdots, \bm{X}_M), \nonumber \\
    &= -\sum_{i,j} \log |\det \bm{W}_i|^2 - \log p(\bm{Y}_1, \bm{Y}_2, \cdots, \bm{Y}_N), \nonumber \\
    &\stackrel{\mathrm{c}}{=} -2J\sum_i |\det \bm{W}_i|
    + \sum_{i,j,n} \left( \frac{ |\bm{w}_{in}^\mathrm{H}\bm{x}_{ij}|^2 }{ \sum_{k} t_{ikn}v_{kjn} } + \log \sum_{k} t_{ikn}v_{kjn} \right), \label{eq:costIlrma}
\end{align}
where $\stackrel{\mathrm{c}}{=}$ denotes equality up to constant factors, and $t_{ikn}>0$ and $v_{kjn}>0$ are the elements of $\bm{T}_n$ and $\bm{V}_n$, respectively.
The minimization of \eqref{eq:costIlrma} can be performed by iterating the following update rules for the spatial model parameters,
\begin{align}
    \bm{U}_{in} &\leftarrow \frac{1}{J} \sum_{j} \frac{1}{\sum_k t_{ikn}v_{kjn}}\bm{x}_{ij}\bm{x}_{ij}^\mathrm{H}, \label{eq:spUpdateU} \\
    \bm{w}_{in} &\leftarrow \left( \bm{W}_i\bm{U}_{in} \right)^{-1}\bm{e}_n, \label{eq:spUpdatew2} \\
    \bm{w}_{in} &\leftarrow \bm{w}_{in} \left( \bm{w}_{in}^\mathrm{H}\bm{U}_{in}\bm{w}_{in} \right)^{-\frac{1}{2}}, \label{eq:spUpdatew3} \\
    y_{ijn} &\leftarrow \bm{w}_{in}^\mathrm{H}\bm{x}_{ij}, \label{eq:sepUpdate}
\end{align}
and for the source model parameters,
\begin{align}
    t_{ikn} &\leftarrow t_{ikn} \sqrt{ \frac{ \sum_j |y_{ijn}|^2 \left( \sum_{k'} t_{ik'n}v_{k'jn} \right)^{-2} v_{kjn} }{ \sum_j \left( \sum_{k'} t_{ik'n}v_{k'jn} \right)^{-1} v_{kjn} } }, \label{eq:srUpdateT} \\
    v_{kjn} &\leftarrow v_{kjn} \sqrt{ \frac{ \sum_i |y_{ijn}|^2 \left( \sum_{k'} t_{ik'n}v_{k'jn} \right)^{-2} t_{ikn} }{ \sum_i \left( \sum_{k'} t_{ik'n}v_{k'jn} \right)^{-1} t_{ikn} } }, \label{eq:srUpdateV}
\end{align}
where $\bm{e}_n\in\{0,1\}^{N}$ is the unit vector with the $n$th element equal to unity.
Update rules \eqref{eq:spUpdateU}--\eqref{eq:srUpdateV} ensure the monotonic non-increase of the negative log-likelihood function $\mathcal{L}$.
After iterative calculations of updates \eqref{eq:spUpdateU}--\eqref{eq:srUpdateV}, the separated signal can be obtained by \eqref{eq:demixingSystem}.

{Equation \eqref{eq:sepUpdate} is equivalent to beamforming~\cite{BrandsteinBook} to $\bm{x}_{ij}$ with the beamformer coefficients $\bm{w}_{in}$. 
Thus, FDICA, IVA, and ILRMA can be interpreted as an adaptive estimation process of beamforming coefficients without having to know the geometry of microphones and sources~\cite{Araki2003}. 
For this reason, the estimated signal $\bm{Y}_{n}$ obtained by \eqref{eq:sepUpdate} is a complex-valued spectrogram, and we do not need to recover its phase components using, e.g., Griffin--Lim-algorithm-based techniques~\cite{YatabeReLU,MasuyamaManifold,MasuyamaADMM,MasuyamaDeGLI,Masuyama2020,Griffin1984,Gunawan2010,Sturmel2012,Watanabe2013,Mayer2017}. 
Both the amplitude and phase components of each source are recovered by the complex-valued linear separation filter $\bm{w}_{in}$.}

\subsection{Proposed Consistent ILRMA}

To further improve the separation performance of the standard ILRMA, we introduce the spectrogram consistency into the parameter update procedure.
In the proposed Consistent ILRMA, the following combination of forward and inverse STFT is performed at the beginning of each iteration {of parameter updates}:
\begin{align}
    \bm{Y}_n \leftarrow \STFT_{\bm{\omega}}( \ISTFT_{\widetilde{\bm{\omega}}}( \bm{Y}_n )). \label{eq:consistUpdate}
\end{align}
This procedure is the projection of the spectrogram of a separated signal $\bm{Y}_n$ onto the set of consistent spectrograms~\cite{Yatabe2020_consistIca}.
That is, $\STFT_{\bm{\omega}}( \ISTFT_{\widetilde{\bm{\omega}}}( \bm{Y}_n ))$ performs nothing if $\bm{Y}_n$ is consistent, but otherwise it smooths the complex spectrogram $\bm{Y}_n$, by going through the time domain, so that the uncertainty principle is satisfied.

{In Consistent ILRMA, the calculation of \eqref{eq:consistUpdate} is performed in each iteration of parameter updates based on \eqref{eq:spUpdateU}--\eqref{eq:srUpdateV}.
Enforcing the spectrogram consistency for the temporary separated signal $\bm{Y}_n$ in each iteration guides the parameters $\bm{W}_i$, $\bm{T}_n$ and $\bm{V}_n$ to better solutions, which results in higher separation performance compared to that of conventional ILRMA.}

Note that this simple update \eqref{eq:consistUpdate} may increase the value of the negative log-likelihood function \eqref{eq:costIlrma}, and therefore the monotonicity of the algorithm is no longer guaranteed. 
However, we will see later in the experiments that the value of the negative log-likelihood function stably decreases as in the standard ILRMA.
The amount of the inconsistent component \eqref{eq:consiste} also settles down to some specific value after several iterations.

\subsection{Iterative back projection}

Since frequency-domain BSS cannot determine the scales of estimated signals (represented by $\bm{D}_i$ in \eqref{eq:indeterminancy}), the spectrogram of a separated signal $\bm{Y}_n$ after an iteration is inconsistent due to the scale irregularity.
To take full advantage of the projection enforcing spectrogram consistency in \eqref{eq:consistUpdate}, we also {propose} applying the following back projection at the end of each iteration so that the frequency-wise scales are aligned.

In determined BSS, the back projection is a standard procedure for recovering the frequency-wise scales.
It can be written as~\cite{Matsuoka2001_PB}:
\begin{equation}
    \tilde{\bm{y}}_{ijn} = \bm{W}_i^{-1} \left( \bm{e}_n \circ \bm{y}_{ij} \right) = y_{ijn}\bm{\lambda}_{in}, \label{eq:backProjection}
\end{equation}
where $\tilde{\bm{y}}_{ijn} = [\, \tilde{y}_{ijn1}, \tilde{y}_{ijn2}, \cdots, \tilde{y}_{ijnM} \,]^\mathrm{T}\in\mathbb{C}^M$ is the $(i,j)$th bin of the scale-fitted spectrogram of the $n$th separated signal, $\bm{\lambda}_{in} = [\,\lambda_{in1}, \lambda_{in2}, \cdots, \lambda_{inM}\,]^\mathrm{T}\in\mathbb{C}^M$ is a coefficient vector of back projection for the $n$th signal at the $i$th frequency, and $\circ$ denotes the element-wise multiplication. 
In the proposed method, this update \eqref{eq:backProjection} is performed at the end of each iteration so that the projection \eqref{eq:consistUpdate} at the beginning of the next iteration properly smooths the spectrograms without the effect of scale indeterminacy.

One side effect of this back projection is that the value of the negative log-likelihood function \eqref{eq:costIlrma} is also changed due to the scale modification.
{In IVA, this problem cannot be avoided because the only parameter in IVA is the demixing matrix $\bm{W}_i$. 
However, in ILRMA, since both the demixing matrix $\bm{W}_i$ and the source model parameter $\bm{T}_n\bm{V}_n$ can determine the scale of estimated signal $\bm{Y}_n$, the likelihood variation can be avoided by appropriately adjusting $\bm{w}_{in}$ and $\bm{T}_n$ after the back projection.}
To prevent the likelihood variation, the following updates are required after performing \eqref{eq:backProjection}:
\begin{align}
    \bm{w}_{in} &\leftarrow \bm{w}_{in} \lambda_{inm_\mathrm{ref}}, \label{eq:normalizew} \\
    y_{ijn} &\leftarrow \bm{w}_{in}^\mathrm{H}\bm{x}_{ij}, \\
    t_{ikn} &\leftarrow t_{ikn} |\lambda_{inm_\mathrm{ref}}|^2, \label{eq:normalizet} 
\end{align}
where $m_\mathrm{ref}$ is the index of the reference channel utilized in the back projection.

The overall algorithm of the proposed Consistent ILRMA is summarized in Algorithm~\ref{alg1}.
{The iterative loop for the parameter optimization appears in the second to eighth lines. 
The spectrogram consistency of the temporary separated signal $\bm{Y}_n$ is ensured in the third line, and the iterative back projection is applied in the sixth and seventh lines. 
Note that an algorithm for the conventional ILRMA can be obtained by performing only the fourth and fifth lines (i.e., ignoring the third, sixth, and seventh lines).
A Python code of the conventional ILRMA is openly available online (\url{https://pyroomacoustics.readthedocs.io/en/pypi-release/pyroomacoustics.bss.ilrma.html}), and therefore the proposed Consistent ILRMA with Python can be easily implemented by slightly modifying the codes.
A MATLAB code of Consistent ILRMA is also available online (\url{https://github.com/d-kitamura/ILRMA/blob/master/consistentILRMA.m}).}

\begin{algorithm}[t!]
    \caption{Consistent ILRMA}
    \label{alg1}
    \setlength{\baselineskip}{13pt}
    \begin{algorithmic}[1]
        \renewcommand{\algorithmicrequire}{\textbf{Input:}}
        \renewcommand{\algorithmicensure}{\textbf{Output:}}
        \REQUIRE $ \{\bm{x}_{ij}\}_{i=1,j=1}^{I,J}, \mathsf{{maxIter}}$
        \ENSURE $\{\bm{y}_{ij}\}_{i=1,j=1}^{I,J}$
        \STATE Initialize $\{\bm{T}_n\}_{n=1}^{N}, \{\bm{V}_n\}_{n=1}^{N}, \{\bm{W}_i\}_{i=1}^{I}$
        \FOR{$\mathsf{iter} = 1, 2, \cdots, \mathsf{maxIter}$}
            \STATE{Ensure consistency by calculating \eqref{eq:consistUpdate}} $\forall n$
            \STATE{Update source model by calculating \eqref{eq:srUpdateT} and \eqref{eq:srUpdateV}} $\forall i,j,k,n$
            \STATE{Update spatial model by calculating \eqref{eq:spUpdateU}--\eqref{eq:sepUpdate}} $\forall i,j,n$
            \STATE{Apply back projection by calculating \eqref{eq:backProjection} $\forall i,j,n$}
            \STATE{Update parameters by calculating \eqref{eq:normalizew}--\eqref{eq:normalizet} $\forall i,j,k,n$}
        \ENDFOR
    \end{algorithmic}
\end{algorithm}

\section{Experiments} \label{sec:experiment}

{In this section, we conducted two experiments using synthesized and real-recorded mixtures.
The synthesized mixtures were produced by convoluting the impulse responses to dry audio sources, while the real-recorded mixtures were actually recorded by using a microphone array in an ordinary room with ambient noise.}

\subsection{BSS of synthesized mixtures} \label{sec:Synth}
\subsubsection{Conditions} \label{sec:SynthCond}

\begin{figure}[t]
    \begin{center}
        \includegraphics[width=0.9\columnwidth]{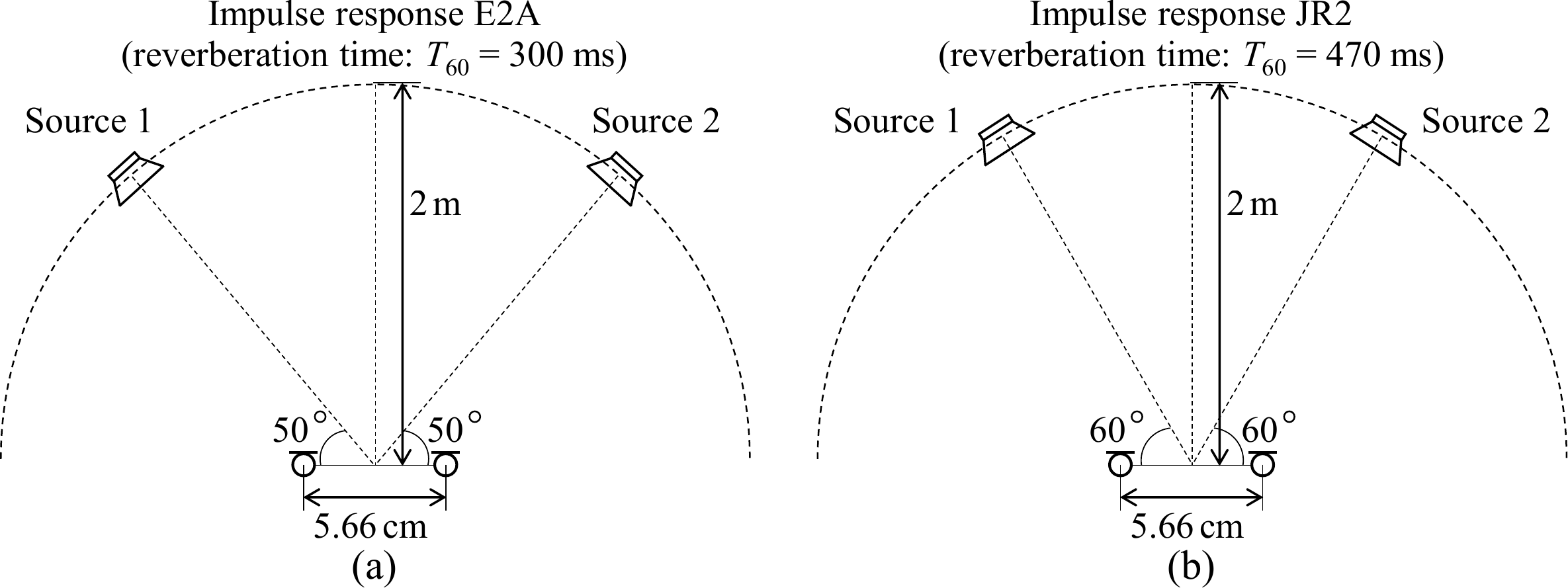}
    \end{center}
    \caption{Recording conditions of impulse responses: (a) E2A and (b) JR2.}
    \label{fig:RIR}
\end{figure}

We conducted determined BSS experiments using {synthesized} music and speech mixtures with two sources and two microphones ($N\!=\!M\!=\!2$).
The dry sources of music and speech signals, listed in Table~\ref{usedSources}, were respectively obtained from \texttt{professionally produced music} and \texttt{underdetermined separation tasks} provided as a part of SiSEC2011~\cite{Araki2012_sisec}.
They were convoluted with the impulse response \texttt{E2A} ($T_{60}=300$\:ms) or \texttt{JR2} ($T_{60}=470$\:ms), obtained from the RWCP database~\cite{Nakamura2000_rwcp}, to simulate the multichannel observation signals.
The recording conditions of these impulse responses are shown in Fig.\:\ref{fig:RIR}.

{In this experiment, we compared the performance of six methods: three conventional and three proposed.
The conventional methods were the standard IVA~\cite{Ono2011_auxiva}, Consistent IVA~\cite{Yatabe2020_consistIca}, and standard ILRMA~\cite{Kitamura2016_ilrma}.
The proposed methods were Consistent IVA with iterative back projection (Consistent IVA+BP), Consistent ILRMA, and Consistent ILRMA with iterative back projection (Consistent ILRMA+BP).
For all methods, the initial demixing matrix was set to an identity matrix. 
For the ILRMA-based methods, the nonnegative matrices $\bm{T}_n$ and $\bm{V}_n$ were initialized using uniformly distributed random values in the range $(0,1)$.}
Five trials were performed for each condition using different pseudorandom seeds.
The number of bases for each source, $K$, was set to $10$ for music mixtures and $2$ for speech mixtures, where it was experimentally confirmed that these conditions provide the best performance for the conventional ILRMA~\cite{Kitamura2016_ilrma}. 
To satisfy the perfect reconstruction condition \eqref{eq:perfectReconstructionCond}, the inverse STFT was implemented by the canonical dual of the analysis window.
{For both Consistent IVA+BP and Consistent ILRMA+BP, the iterative back projection was applied,} where the reference channel was set to $m_\mathrm{ref}\!=\!1$.
Since the property of spectrogram consistency depends on the window length, shift length, {and type of window function,} various combinations of them were tested. 
The experimental conditions are summarized in Table~\ref{conditions}.

\begingroup
 \renewcommand{\arraystretch}{1}
\begin{table}[t]
\vspace{10pt}
\caption{Music and speech dry sources obtained from SiSEC2011}
\label{usedSources}
\begin{center}
\begin{tabular}{ccc}
\Hline
 \raisebox{-0.2ex}[0cm][0cm]{Signal}		& \raisebox{-0.2ex}[0cm][0cm]{Data name}									& \raisebox{-0.2ex}[0cm][0cm]{Source (1\texttt{/}2)} \\ \hline
 \raisebox{-0.2ex}[0cm][0cm]{Music 1}		& \raisebox{-0.2ex}[0cm][0cm]{bearlin-roads}								& \raisebox{-0.2ex}[0cm][0cm]{acoustic\_guit\_main\texttt{/}vocals} \\ \hline
 \raisebox{-0.2ex}[0cm][0cm]{{Music 2}}	& \raisebox{-0.2ex}[0cm][0cm]{{bearlin-roads}}							& \raisebox{-0.2ex}[0cm][0cm]{{piano\texttt{/}acoustic\_guit\_main}} \\ \hline
 \raisebox{-0.2ex}[0cm][0cm]{{Music 3}}	& \raisebox{-0.2ex}[0cm][0cm]{{bearlin-roads}}							& \raisebox{-0.2ex}[0cm][0cm]{{piano\texttt{/}vocals}} \\ \hline
 \raisebox{-0.2ex}[0cm][0cm]{Music 4}		& \raisebox{-0.2ex}[0cm][0cm]{another\_dreamer-the\_ones\_we\_love}			& \raisebox{-0.2ex}[0cm][0cm]{guitar\texttt{/}vocals} \\ \hline
 \raisebox{-0.2ex}[0cm][0cm]{{Music 5}}	& \raisebox{-0.2ex}[0cm][0cm]{{another\_dreamer-the\_ones\_we\_love}}	& \raisebox{-0.2ex}[0cm][0cm]{{drums\texttt{/}guitar}} \\ \hline
 \raisebox{-0.2ex}[0cm][0cm]{Music 6}		& \raisebox{-0.2ex}[0cm][0cm]{fort\_minor-remember\_the\_name}				& \raisebox{-0.2ex}[0cm][0cm]{violins\_synth\texttt{/}vocals} \\ \hline
 \raisebox{-0.2ex}[0cm][0cm]{{Music 7}}	& \raisebox{-0.2ex}[0cm][0cm]{{fort\_minor-remember\_the\_name}}			& \raisebox{-0.2ex}[0cm][0cm]{{vocals\texttt{/}drums}} \\ \hline
 \raisebox{-0.2ex}[0cm][0cm]{{Music 8}}	& \raisebox{-0.2ex}[0cm][0cm]{{tamy-que\_pena\_tanto\_faz}}				& \raisebox{-0.2ex}[0cm][0cm]{{guitar\texttt{/}vocals}} \\ \hline
 \raisebox{-0.2ex}[0cm][0cm]{Music 9}		& \raisebox{-0.2ex}[0cm][0cm]{ultimate\_nz\_tour}							& \raisebox{-0.2ex}[0cm][0cm]{guitar\texttt{/}synth} \\ \hline
 \raisebox{-0.2ex}[0cm][0cm]{{Music 10}}	& \raisebox{-0.2ex}[0cm][0cm]{{ultimate\_nz\_tour}}						& \raisebox{-0.2ex}[0cm][0cm]{{drums\texttt{/}vocals}} \\ \hline
 \raisebox{-0.2ex}[0cm][0cm]{Speech 1}		& \raisebox{-0.2ex}[0cm][0cm]{dev1\_female4}								& \raisebox{-0.2ex}[0cm][0cm]{src\_1\texttt{/}src\_2} \\ \hline
 \raisebox{-0.2ex}[0cm][0cm]{{Speech 2}}	& \raisebox{-0.2ex}[0cm][0cm]{{dev1\_female4}}							& \raisebox{-0.2ex}[0cm][0cm]{{src\_1\texttt{/}src\_4}} \\ \hline
 \raisebox{-0.2ex}[0cm][0cm]{{Speech 3}}	& \raisebox{-0.2ex}[0cm][0cm]{{dev1\_female4}}							& \raisebox{-0.2ex}[0cm][0cm]{{src\_2\texttt{/}src\_3}} \\ \hline
 \raisebox{-0.2ex}[0cm][0cm]{{Speech 4}}	& \raisebox{-0.2ex}[0cm][0cm]{{dev1\_female4}}							& \raisebox{-0.2ex}[0cm][0cm]{{src\_2\texttt{/}src\_4}} \\ \hline
 \raisebox{-0.2ex}[0cm][0cm]{Speech 5}		& \raisebox{-0.2ex}[0cm][0cm]{dev1\_female4}								& \raisebox{-0.2ex}[0cm][0cm]{src\_3\texttt{/}src\_4} \\ \hline
 \raisebox{-0.2ex}[0cm][0cm]{Speech 6}		& \raisebox{-0.2ex}[0cm][0cm]{dev1\_male4}									& \raisebox{-0.2ex}[0cm][0cm]{src\_1\texttt{/}src\_2} \\ \hline
 \raisebox{-0.2ex}[0cm][0cm]{{Speech 7}}	& \raisebox{-0.2ex}[0cm][0cm]{{dev1\_male4}}								& \raisebox{-0.2ex}[0cm][0cm]{{src\_1\texttt{/}src\_4}} \\ \hline
 \raisebox{-0.2ex}[0cm][0cm]{{Speech 8}}	& \raisebox{-0.2ex}[0cm][0cm]{{dev1\_male4}}								& \raisebox{-0.2ex}[0cm][0cm]{{src\_2\texttt{/}src\_3}} \\ \hline
 \raisebox{-0.2ex}[0cm][0cm]{{Speech 9}}	& \raisebox{-0.2ex}[0cm][0cm]{{dev1\_male4}}								& \raisebox{-0.2ex}[0cm][0cm]{{src\_2\texttt{/}src\_4}} \\ \hline
 \raisebox{-0.2ex}[0cm][0cm]{Speech 10}		& \raisebox{-0.2ex}[0cm][0cm]{dev1\_male4}									& \raisebox{-0.2ex}[0cm][0cm]{src\_3\texttt{/}src\_4} \\ 
\Hline
\end{tabular}
\end{center}
\end{table}
\endgroup

\begingroup
 \renewcommand{\arraystretch}{1}
\begin{table}[t]
\caption{Experimental conditions}
\label{conditions}
\begin{center}
\begin{tabular}{cc}
\Hline
 \raisebox{-0.2ex}[0cm][0cm]{Window function}		& \raisebox{-0.2ex}[0cm][0cm]{Hann/{Hamming/Blackman} window}	 \\ \hline
 \raisebox{-0.2ex}[0cm][0cm]{Window length}			& \raisebox{-0.2ex}[0cm][0cm]{64, 128, 256, 512, 768, 1024~ms} \\ \hline
 \raisebox{-0.2ex}[0cm][0cm]{Window shift length}	& \raisebox{-0.2ex}[0cm][0cm]{1/16, 1/8, 1/4, 1/2 of window length} \\ \hline
 \raisebox{-0.2ex}[0cm][0cm]{Number of bases $K$ for}	& \raisebox{-0.2ex}[0cm][0cm]{10 for music signals} \\
 \raisebox{-0.2ex}[0cm][0cm]{each source in ILRMA}	& \raisebox{-0.2ex}[0cm][0cm]{and 2 for speech signals} \\ \hline
 \raisebox{-0.2ex}[0cm][0cm]{Number of iterations}	& \raisebox{-0.2ex}[0cm][0cm]{100} \\ 
\Hline
\end{tabular}
\end{center}
\end{table}
\endgroup

{For quantitative evaluation of the separation performance, we measured the source-to-distortion ratio (SDR), source-to-interference ratio (SIR), and sources-to-artifact ratio (SAR).
In a noiseless situation, SDR, SIR, and SAR are defined as follows~\cite{Vincent2006_bssEval}:
\begin{align}
    \mathrm{SDR} &= 10\log_{10} \frac{ \sum_{l} |s_\mathrm{t}[l]|^2 }{ \sum_{l} |e_\mathrm{i}[l] + e_\mathrm{a}[l] |^2 }, \\
    \mathrm{SIR} &= 10\log_{10} \frac{ \sum_{l} |s_\mathrm{t}[l]|^2 }{ \sum_{l} |e_\mathrm{i}[l]|^2 }, \\
    \mathrm{SAR} &= 10\log_{10} \frac{ \sum_{l} |s_\mathrm{t}[l]+e_\mathrm{i}[l]|^2 }{ \sum_{l} |e_\mathrm{a}[l]|^2 },
\end{align}
where $s_\mathrm{t}[l]$, $e_\mathrm{i}[l]$, and $e_\mathrm{a}[l]$ are the $l$th sample of target signal, interference, and artificial components of the estimated signal, respectively, in the time domain. 
SIR and SAR are used to quantify the amount of interference rejection and the absence of artificial distortion of the estimated signal, respectively.
SDR is used to quantify the overall separation performance, as SDR is in good agreement with both SIR and SAR for determined BSS.}

{In this experiment, the energy of sources was not adjusted, i.e., the energy ratio of sources (source-to-source ratio) was automatically determined by the initial volume of the dry sources and the level of the impulse responses.
That is, the source-to-source ratio of each mixture signal is different from the others.
To equally evaluate the performances of different mixtures, we calculated SDR improvement ($\Delta$SDR) and SIR improvement ($\Delta$SIR) defined as
\begin{align}
    \Delta\mathrm{SDR} &= \mathrm{SDR}_\mathrm{sep} - \mathrm{SDR}_\mathrm{input}, \\
	\Delta\mathrm{SIR} &= \mathrm{SIR}_\mathrm{sep} - \mathrm{SIR}_\mathrm{input},
\end{align}
where $\mathrm{SDR}_\mathrm{sep}$ and $\mathrm{SIR}_\mathrm{sep}$ are the SDR and SIR of the separated signal, and $\mathrm{SDR}_\mathrm{input}$ and $\mathrm{SIR}_\mathrm{input}$ are the SDR and SIR of the initial mixture signal input to the BSS methods. 
Note that SAR improvement cannot be defined because its value of the signal without artificial processing cannot be defined ($\mathrm{SAR}_\mathrm{input}=\infty$).}

\subsubsection{Results and discussions} \label{sec:SynthResults}

\begin{figure}[t]
    \begin{center}
        \includegraphics[width=0.5\columnwidth]{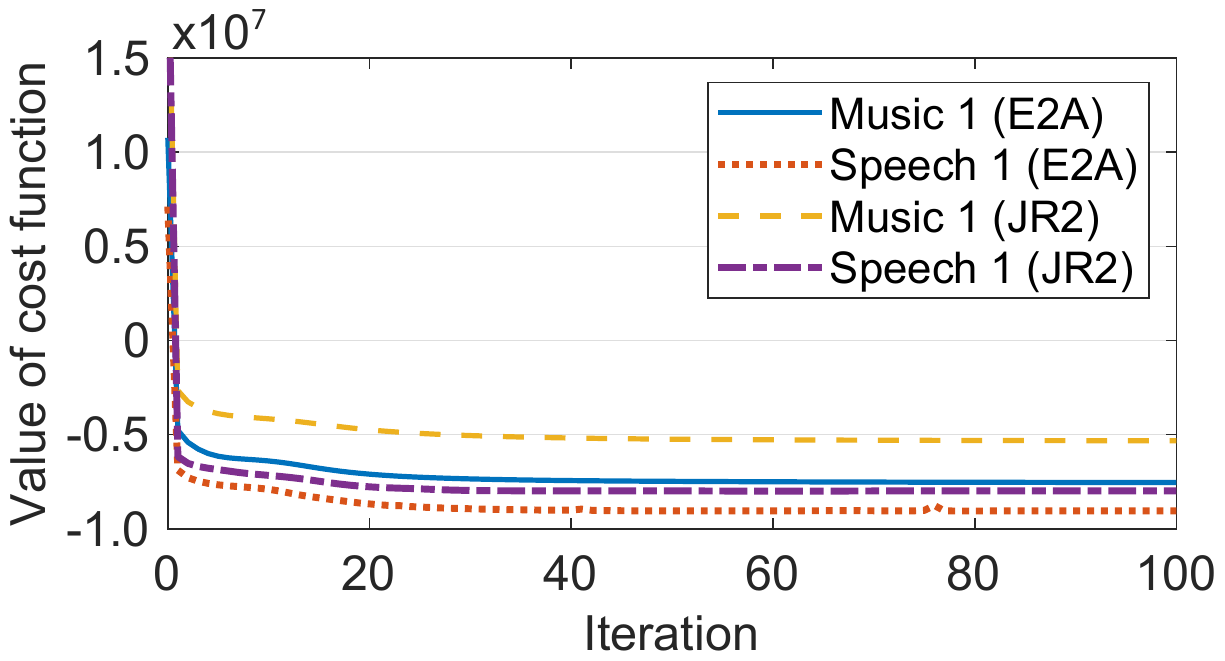}
    \end{center}
    \caption{Values of negative log-likelihood function \eqref{eq:costIlrma} of {Consistent ILRMA+BP} (window length: 256\:ms, shift length: 32\:ms).}
    \label{fig:costCurve}
\end{figure}

\begin{figure}[t]
    \centering
    \subfloat[Window length: 256~ms, shift length: 32~ms]{
        \includegraphics[width=0.5\columnwidth]{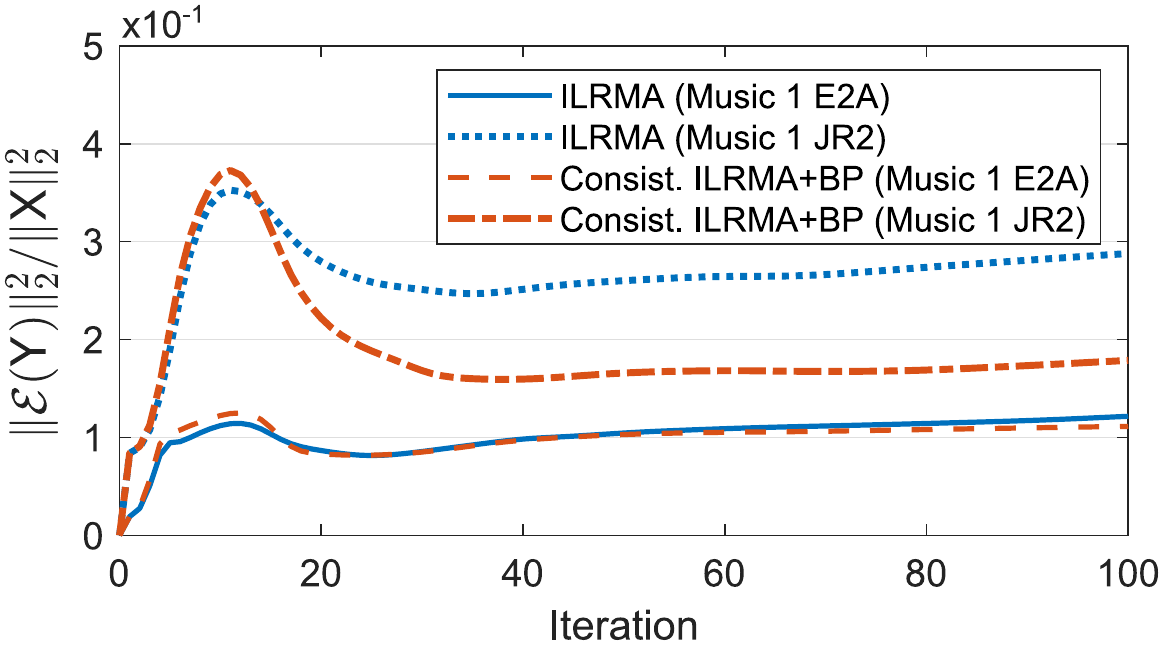}} \\
    \subfloat[Window length: 1024~ms, shift length: 512~ms]{
        \includegraphics[width=0.5\columnwidth]{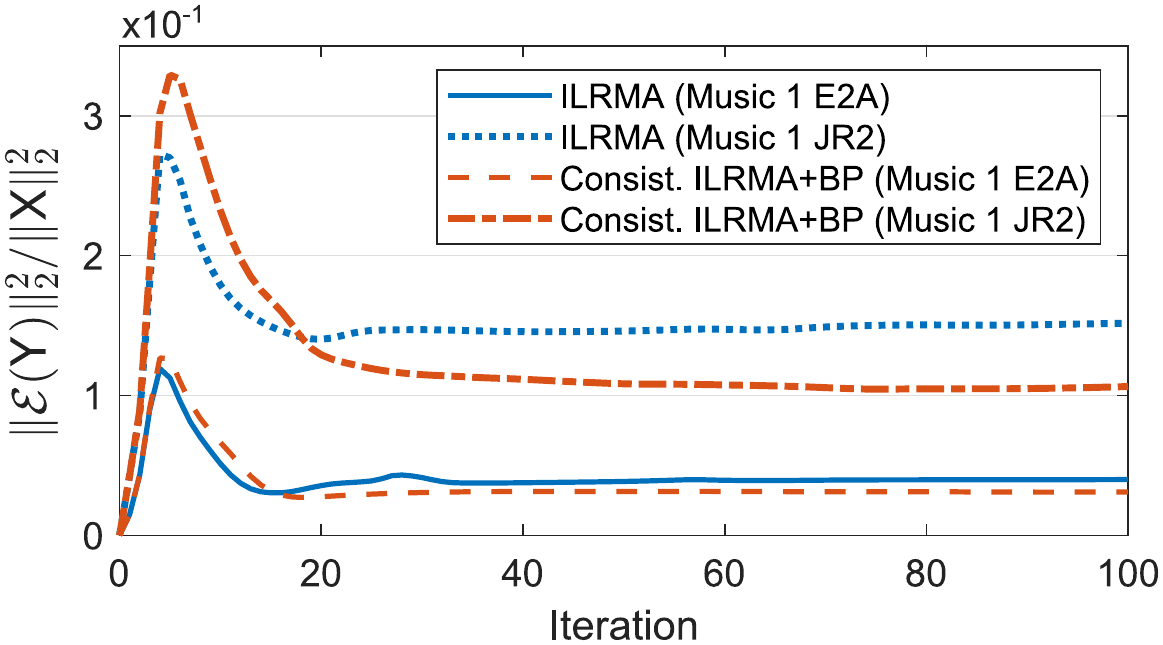}}
    \caption{{Examples of normalized energy of inconsistent components $(\|\mathcal{E}(\mathsf{Y})\|_2^2 / \|\mathsf{X}\|_2^2)$ of ILRMA and Consistent ILRMA+BP for Music 1: (a) 256-ms-long window and 32-ms shifting and (b) 1024-ms-long window and 512-ms shifting, where $\mathsf{X}=[\bm{X}_1, \bm{X}_2]$, and $\mathcal{E}(\cdot)$ is in \eqref{eq:consiste}.}}
    \label{fig:diffCurveMus}
\end{figure}

\begin{figure}[t]
    \centering
    \subfloat[Window length: 256~ms, shift length: 32~ms]{
        \includegraphics[width=0.5\columnwidth]{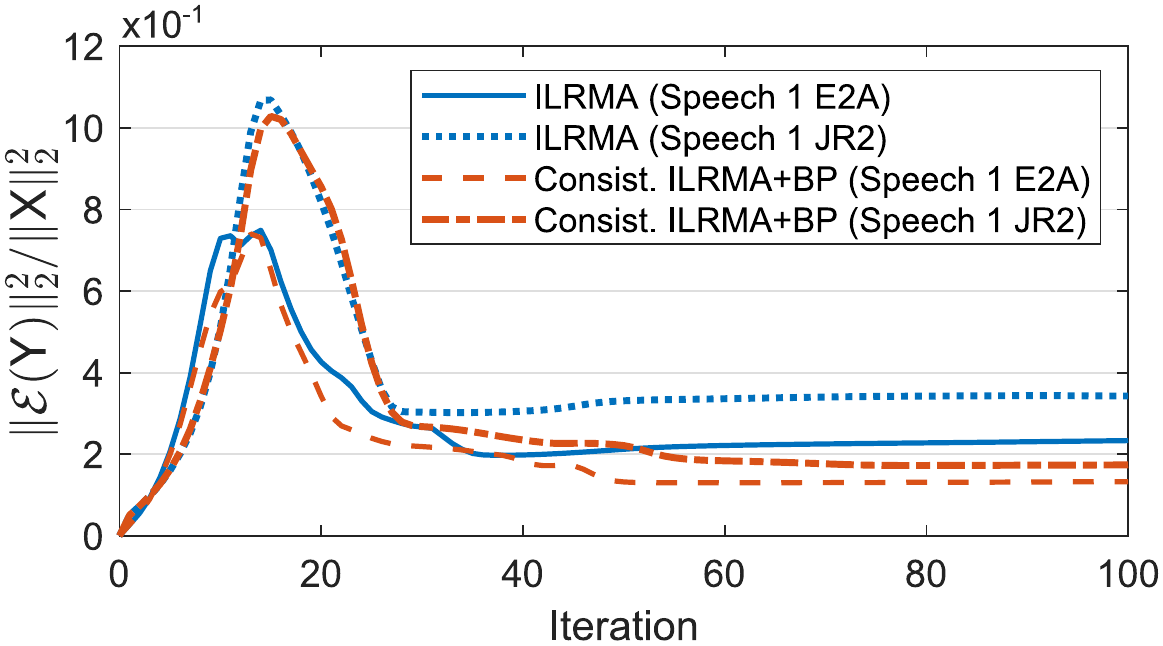}} \\
    \subfloat[Window length: 1024~ms, shift length: 512~ms]{
        \includegraphics[width=0.5\columnwidth]{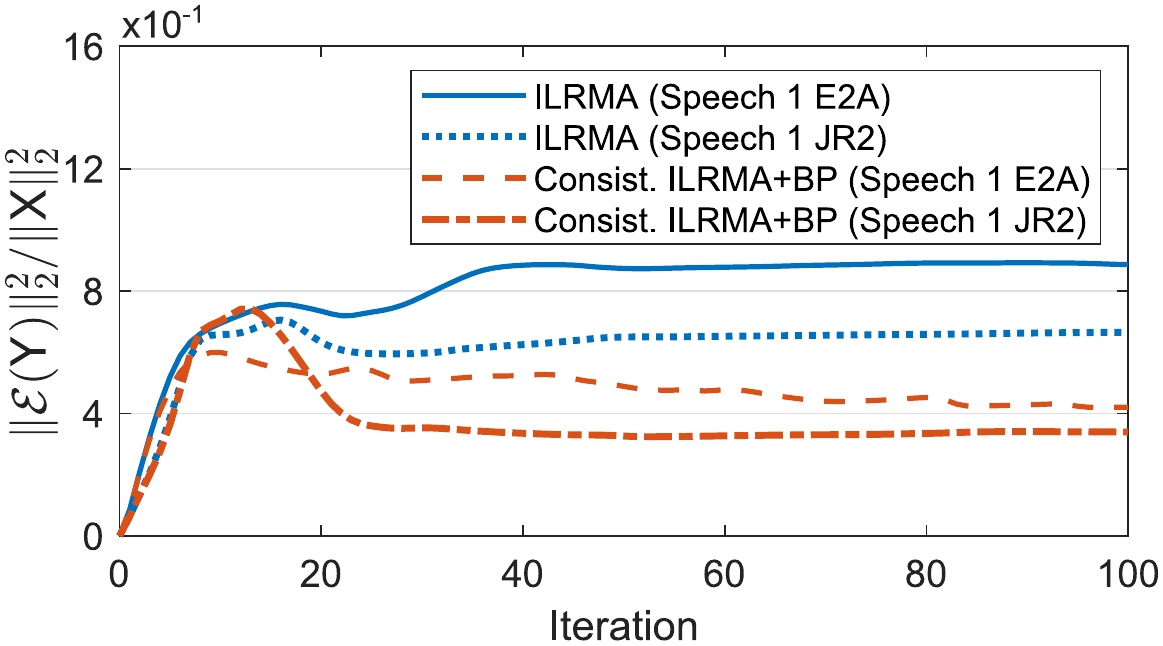}} \\
    \caption{{Examples of normalized energy of inconsistent components $(\|\mathcal{E}(\mathsf{Y})\|_2^2 / \|\mathsf{X}\|_2^2)$ of ILRMA and Consistent ILRMA+BP for Speech 1: (a) 256-ms-long window and 32-ms shifting and (b) 1024-ms-long window and 512-ms shifting, where $\mathsf{X}=[\bm{X}_1, \bm{X}_2]$, and $\mathcal{E}(\cdot)$ is in \eqref{eq:consiste}.}}
    \label{fig:diffCurveSpe}
\end{figure}

Figure\:\ref{fig:costCurve} shows examples of the value of the negative log-likelihood function \eqref{eq:costIlrma} of {Consistent ILRMA+BP.}
Although the algorithmic convergence of the proposed method has not been theoretically justified because of the additional projection \eqref{eq:consistUpdate}, we experimentally confirmed a smooth decrease of the cost function.
We also confirmed that such behavior was common for the other experimental conditions and mixtures.
This result indicates that the additional procedure in the proposed method does not have a harmful effect on the behavior of the overall algorithm.

Figures\:\ref{fig:diffCurveMus} and \ref{fig:diffCurveSpe} show examples of the energy of the inconsistent components \eqref{eq:consiste} of {standard ILRMA and Consistent ILRMA+BP}. 
The energy was normalized by that of the initial spectrograms in order to align the vertical axis.
{Note that the energy of inconsistency components is not directly related to the degree of permutation misalignment or the separation performance.
These figures are shown to confirm whether the proposed algorithm can properly reduce the degree of inconsistency.}
These values are completely zero when the separated spectrograms are consistent, and hence those at the 0th iteration (the leftmost values) are zero because no processing is performed at that point.
By iterating {the algorithms}, this energy rapidly increased because the demixing matrix for each frequency independently tried to process and separate the signals.
However, the normalized energy tended toward some specific values after several iterations. 
{We confirmed that the converged values of Consistent ILRMA+BP were always lower than those of standard ILRMA.}
This result indicates that {Consistent ILRMA+BP} reduces the amount of the inconsistent components and tries to make the separated spectrogram more consistent.
{In addition, similar to Fig.\:\ref{fig:costCurve}, the algorithmic stability of Consistent ILRMA+BP can be confirmed from Figs.\:\ref{fig:diffCurveMus} and \ref{fig:diffCurveSpe}.}

\begin{figure*}[p]
    \begin{center}
        \includegraphics[width=0.96\columnwidth]{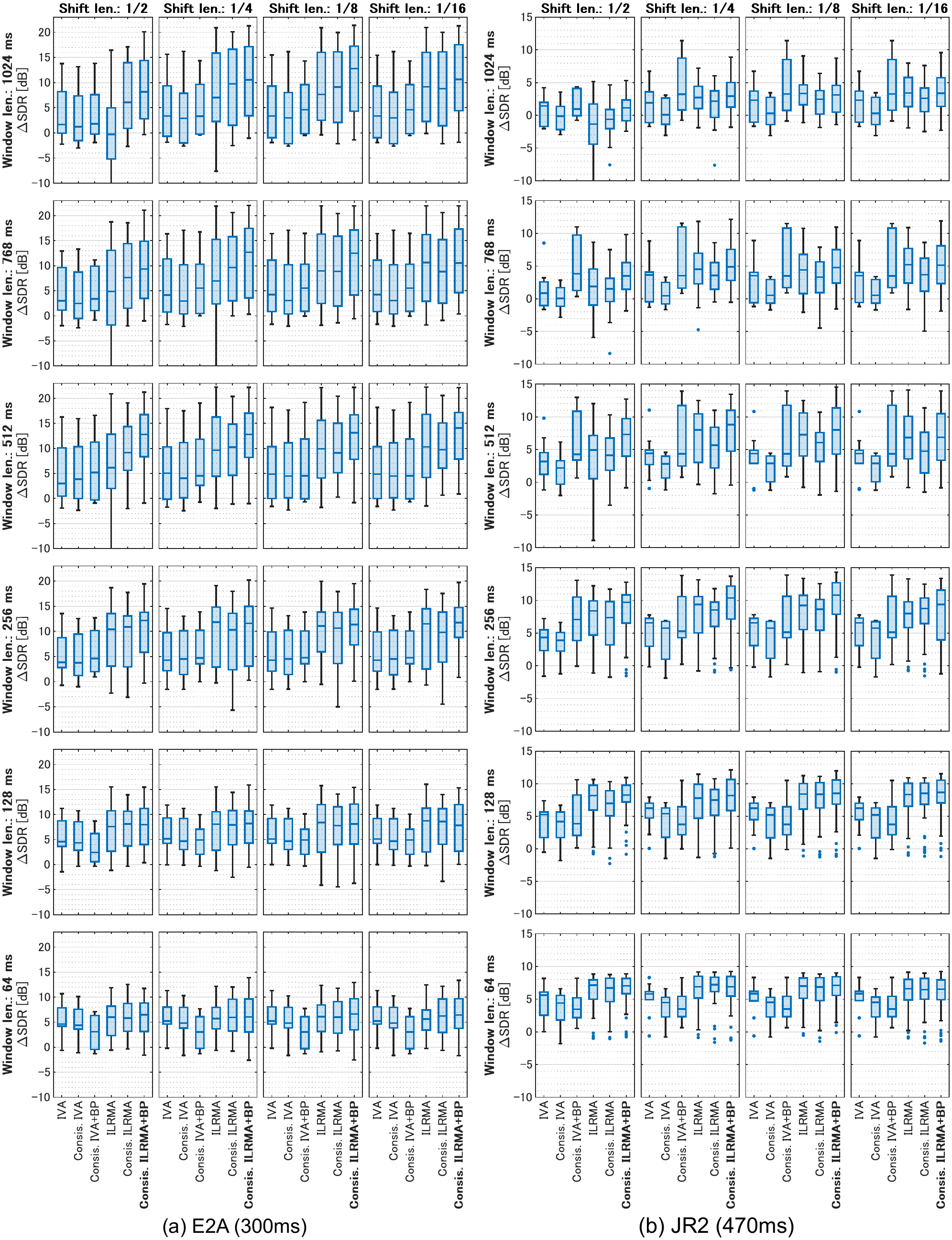}
    \end{center}
    \caption{Average SDR improvements for synthesized music mixtures (Music 1--10) with (a) \texttt{E2A} and (b) \texttt{JR2}, where Hann window is used in STFT.}
    \label{fig:sdr_Music_hann}
\end{figure*}

\begin{figure*}[p]
    \begin{center}
        \includegraphics[width=0.96\columnwidth]{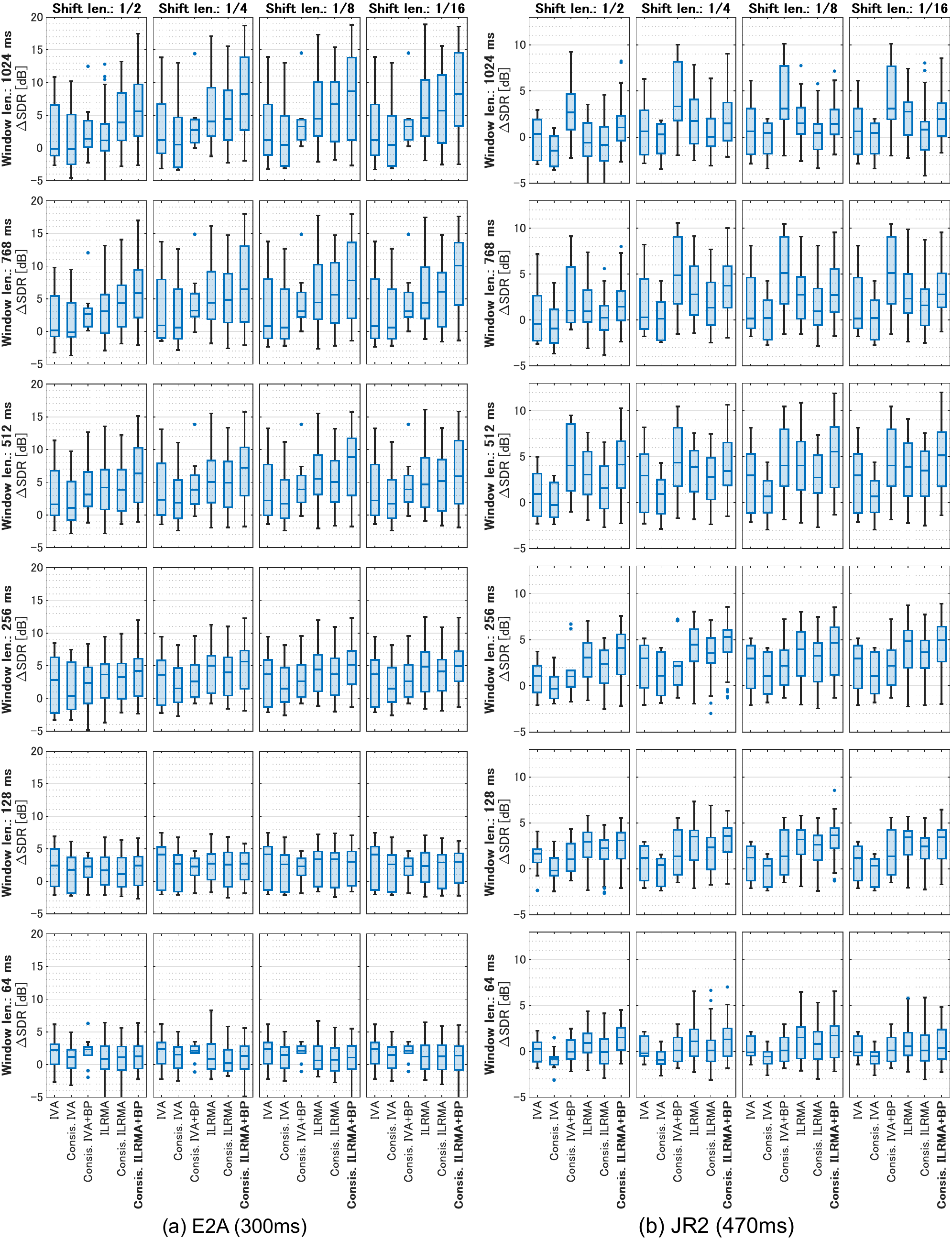}
    \end{center}
    \caption{Average SDR improvements for synthesized speech mixtures (Speech 1--10) with (a) \texttt{E2A} and (b) \texttt{JR2}, where Hann window is used in STFT.}
    \label{fig:sdr_Speech_hann}
\end{figure*}

{Figures\:\ref{fig:sdr_Music_hann} and \ref{fig:sdr_Speech_hann} summarize the SDR improvements for the music mixtures and speech mixtures, respectively.
The window function was the Hann window.
Each box contains 50 results (i.e., 5 pseudorandom seeds $\times$ 10 mixtures in Table~\ref{usedSources}), where $\Delta$SDRs of the two separated sources in each mixture were averaged.
The central lines of the box plots indicate the median, and the bottom and top edges of the box indicate the 25th and 75th percentiles, respectively.
Each row corresponds to the same window length, while each column corresponds to the same shift length.
As we conducted the experiment for six window lengths, four shift lengths, and two impulse responses, each figure consists of $6\times4\times2$ subfigures.
In each subfigure, six boxes are shown to illustrate the results of (1) IVA, (2) Consistent IVA, (3) Consistent IVA+BP, (4) ILRMA, (5) Consistent ILRMA, and (6) Consistent ILRMA+BP.
Since the tendency of the results was the same as Figs.\:\ref{fig:sdr_Music_hann} and \ref{fig:sdr_Speech_hann}, we provide the SDR improvements for the other windows (Hamming and Blackman) in the Appendix.
The SIR improvement and SAR are also given in the Appendix.

Since IVA and ILRMA assume the instantaneous mixing model \eqref{eq:mixingSystem} for each frequency in the time-frequency domain, the window length should be long relative to the reverberation time to achieve accurate separation.
At the same time, too long a window degrades the separation performance of IVA and ILRMA, as discussed in~\cite{Kitamura2017_ilrma}. 
This is because capturing the source activity and spectral patterns becomes difficult for IVA and ILRMA as the time resolution of the spectrograms becomes low due to a long window.
The robustness of IVA and ILRMA is also deteriorated by a long window because the effective number of time segments is decreased.
This trade-off of the separation performance caused by window length in STFT can be easily confirmed from the results for both music (Fig.\:\ref{fig:sdr_Music_hann}) and speech (Fig.\:\ref{fig:sdr_Speech_hann}) mixtures, which is consistent with the results in~\cite{Kitamura2017_ilrma}.
As shown in the figures, the performance was poor for the shorter windows ($\leq 128$~ms), and the performance for the longer windows ($\geq 768$~ms) was more varied than that of the shorter ones.
The window length best suited for these conditions (combinations of source signals and impulse responses) seems to be around 256~ms or 512~ms.
While the maximum achievable performance becomes higher as the window length becomes longer due to the mixing model \eqref{eq:mixingSystem}, these results indicate that the source modeling becomes difficult for both IVA and ILRMA when the window length is too long.
This trade-off should be important for discussing the results further. 

By comparing the performances of the conventional (IVA, Consistent IVA, and ILRMA) and proposed (Consistent IVA+BP, Consistent ILRMA, and Consistent ILRMA+BP) methods, we can see that the proposed methods tend to outperform the conventional ones.
Some comparisons are made as follows.
\begin{description}
    \item[\;\;\textbullet\;\;Conventional and proposed IVAs:]
    The proposed Consistent IVA+BP performed better than the conventional IVAs (IVA and Consistent IVA) in Figs.\:\ref{fig:sdr_Music_hann}(b) and \ref{fig:sdr_Speech_hann}(b) when the window length was sufficiently long ($\geq$ 256\:ms).
    In those cases, the conventional Consistent IVA resulted in a worse performance than IVA, which indicates that just using spectrogram consistency cannot improve the performance of IVA.
    This demonstrates the importance of the iterative back projection when spectrogram consistency is considered within determined BSS.
    \item[\;\;\textbullet\;\;Conventional and proposed ILRMAs:]
    The proposed Consistent ILRMA without BP performed comparably to the conventional ILRMA.
    In Figs.\:\ref{fig:sdr_Music_hann}(a) and \ref{fig:sdr_Speech_hann}(a), Consistent ILRMA performed better than ILRMA when the window length was long ($\geq$ 768\:ms).
    In contrast, in Figs.\:\ref{fig:sdr_Music_hann}(b) and \ref{fig:sdr_Speech_hann}(b), Consistent ILRMA performed worse than ILRMA.
    This is presumably because the scale ambiguity prevented the spectrogram consistency from working properly.
    By incorporating iterative back projection into Consistent ILRMA, the proposed Consistent ILRMA+BP performed better than the conventional ILRMA.
    In the best situation (the top left subfigure of Fig.\:\ref{fig:sdr_Music_hann}), Consistent ILRMA+BP performed 8\:dB better than ILRMA by bringing out the potential of spectrogram consistency in determined BSS.
\end{description}

To further explain the experimental results, some notable tendencies are summarized as follows.
\begin{description}
    \item[\;\;\textbullet\;\;Short window:]
    When the window length was short (64\:ms), all methods performed similarly in terms of $\Delta$SDR.
    This is because the achievable performance was already limited by the window length that was shorter than the reverberation time.
    This result contradicted our expectation before performing the experiment.
    Since enforcing the consistency spreads the frequency components based on the main-lobe of the window function, we expected that the ability to solve the permutation problem would be higher when the window length was shorter because of the wider main-lobe.
    In reality, we found that the spectrogram consistency could assist IVA and ILRMA except for the cases where the window length was short ($\leq 128$~ms in this experiment) compared to the reverberation time.
    \item[\;\;\textbullet\;\;Large window shift:]
    When the shift length was 1/2 of the window length, the performance of ILRMA significantly dropped compared to smaller shift lengths ($1/4$, $1/8$ and $1/16$), especially when the window length was long (e.g., 1024\:ms).
    This is presumably because the number of time segments was small, i.e., NMF in ILRMA failed to model the source signals from the given amount of data.
    In addition, for a larger window, distinguishing spectral patterns of the sources became difficult for ILRMA due to the time-directional blurring effect caused by a longer window.
    Such performance degradation was alleviated for Consistent ILRMA+BP.
    This might be because the smoothing process of the inverse STFT provides some additional information for the source modeling from the adjacent bins. 
    \item[\;\;\textbullet\;\;Length of boxes:]
    When the length of the box of ILRMA was long, as in Figs.\:\ref{fig:sdr_Music_hann}(a) and \ref{fig:sdr_Speech_hann}(a), Consistent ILRMA+BP was able to improve the performance.
    Conversely, when the length of the box of ILRMA was short, as in Figs.\:\ref{fig:sdr_Music_hann}(b) and \ref{fig:sdr_Speech_hann}(b), Consistent ILRMA+BP was only able to slightly improve the performance.
    Note that the vertical axes are different.
    This result indicates that the achievable performance decided by the mixing model \eqref{eq:mixingSystem} limits the improvement obtained by spectrogram consistency.
    Since consistency is the characteristic of a spectrogram, it cannot manage the mixing process.
    The demixing-filter update of ILRMA, which is the same for the conventional and proposed methods, manages the mixing process.
    Hence, when the mixing model has a mismatch with the observed condition, there is less room for spectrogram consistency to improve the performance.
    \item[\;\;\textbullet\;\;Improvement by consistency:]
    The proposed method tended to achieve a good performance when the conventional ILRMA also worked well, e.g., Figs.\:\ref{fig:sdr_Music_hann}(a) and \ref{fig:sdr_Music_hann}(a).
    This tendency indicates that the spectrogram consistency effectively promotes the separation when the estimated source $\bm{Y}_n$ accurately approaches the original source $\bm{S}_n$ during the optimization, as $\bm{S}_n$ is naturally a consistent spectrogram.
    This is the reason we feel that the consistency can be an assistant of the frequency-domain BSS.
    An important aspect is that the source model (e.g., NMF in ILRMA) actually informs the separation cue, and the spectrogram consistency enhances the separation performance when the source modeling functions correctly.
\end{description}
}

\subsection{{BSS of real-recorded mixtures}}
\subsubsection{{Conditions}}

{Next, we evaluated the conventional and proposed methods using live-recorded music and speech mixtures obtained from \texttt{underdetermined separation tasks} in SiSEC2011~\cite{Araki2012_sisec}, where only two sources were mixed to make the BSS problem determined ($M=N=2$). 
The signals used in this experiment are listed in Table~\ref{liveRecSources}. 
The reverberation time of these signals was 250~ms, and the microphone spacing was 1~m (see~\cite{Araki2012_sisec}). 
Since these source signals were actually recorded using a microphone array in an ordinary room with ambient noise, the observed signals are more realistic compared to those in Sect.~\ref{sec:Synth}.

For simplicity, in this experiment we used STFT with a fixed condition, the 512-ms-long Hann window with 1/4 shifting.
The experimental conditions other than the window were the same as those in Sect.~\ref{sec:SynthCond}.}

\begingroup
 \renewcommand{\arraystretch}{1}
\begin{table}[t]
\vspace{10pt}
\caption{{Live-recorded music and speech signals obtained from SiSEC2011}}
\label{liveRecSources}
\begin{center}
\begin{tabular}{cc}
\Hline
 \raisebox{-0.2ex}[0cm][0cm]{Signal}	& \raisebox{-0.2ex}[0cm][0cm]{Source (1\texttt{/}2)} \\ \hline
 \raisebox{-0.2ex}[0cm][0cm]{Music 1}	& \raisebox{-0.2ex}[0cm][0cm]{dev1\_nodrums\_liverec\_250ms\_1m\_sim\_1\texttt{/}dev1\_nodrums\_liverec\_250ms\_1m\_sim\_2} \\ \hline
 \raisebox{-0.2ex}[0cm][0cm]{Music 2}	& \raisebox{-0.2ex}[0cm][0cm]{dev1\_nodrums\_liverec\_250ms\_1m\_sim\_1\texttt{/}dev1\_nodrums\_liverec\_250ms\_1m\_sim\_3} \\ \hline
 \raisebox{-0.2ex}[0cm][0cm]{Music 3}	& \raisebox{-0.2ex}[0cm][0cm]{dev1\_nodrums\_liverec\_250ms\_1m\_sim\_2\texttt{/}dev1\_nodrums\_liverec\_250ms\_1m\_sim\_3} \\ \hline
 \raisebox{-0.2ex}[0cm][0cm]{Music 4}	& \raisebox{-0.2ex}[0cm][0cm]{dev1\_wdrums\_liverec\_250ms\_1m\_sim\_1\texttt{/}dev1\_nodrums\_liverec\_250ms\_1m\_sim\_2} \\ \hline
 \raisebox{-0.2ex}[0cm][0cm]{Music 5}	& \raisebox{-0.2ex}[0cm][0cm]{dev1\_wdrums\_liverec\_250ms\_1m\_sim\_1\texttt{/}dev1\_nodrums\_liverec\_250ms\_1m\_sim\_3} \\ \hline
 \raisebox{-0.2ex}[0cm][0cm]{Music 6}	& \raisebox{-0.2ex}[0cm][0cm]{dev1\_wdrums\_liverec\_250ms\_1m\_sim\_2\texttt{/}dev1\_nodrums\_liverec\_250ms\_1m\_sim\_3} \\ \hline
 \raisebox{-0.2ex}[0cm][0cm]{Music 7}	& \raisebox{-0.2ex}[0cm][0cm]{dev2\_nodrums\_liverec\_250ms\_1m\_sim\_1\texttt{/}dev1\_nodrums\_liverec\_250ms\_1m\_sim\_2} \\ \hline
 \raisebox{-0.2ex}[0cm][0cm]{Music 8}	& \raisebox{-0.2ex}[0cm][0cm]{dev2\_nodrums\_liverec\_250ms\_1m\_sim\_1\texttt{/}dev1\_nodrums\_liverec\_250ms\_1m\_sim\_3} \\ \hline
 \raisebox{-0.2ex}[0cm][0cm]{Music 9}	& \raisebox{-0.2ex}[0cm][0cm]{dev2\_nodrums\_liverec\_250ms\_1m\_sim\_2\texttt{/}dev1\_nodrums\_liverec\_250ms\_1m\_sim\_3} \\ \hline
 \raisebox{-0.2ex}[0cm][0cm]{Music 10}	& \raisebox{-0.2ex}[0cm][0cm]{dev2\_wdrums\_liverec\_250ms\_1m\_sim\_1\texttt{/}dev1\_nodrums\_liverec\_250ms\_1m\_sim\_2} \\ \hline
 \raisebox{-0.2ex}[0cm][0cm]{Music 11}	& \raisebox{-0.2ex}[0cm][0cm]{dev2\_wdrums\_liverec\_250ms\_1m\_sim\_1\texttt{/}dev1\_nodrums\_liverec\_250ms\_1m\_sim\_3} \\ \hline
 \raisebox{-0.2ex}[0cm][0cm]{Music 12}	& \raisebox{-0.2ex}[0cm][0cm]{dev2\_wdrums\_liverec\_250ms\_1m\_sim\_2\texttt{/}dev1\_nodrums\_liverec\_250ms\_1m\_sim\_3} \\ \hline
 \raisebox{-0.2ex}[0cm][0cm]{Speech 1}	& \raisebox{-0.2ex}[0cm][0cm]{dev1\_female4\_liverec\_250ms\_1m\_sim\_1\texttt{/}dev1\_female4\_liverec\_250ms\_1m\_sim\_2} \\ \hline
 \raisebox{-0.2ex}[0cm][0cm]{Speech 2}	& \raisebox{-0.2ex}[0cm][0cm]{dev1\_female4\_liverec\_250ms\_1m\_sim\_3\texttt{/}dev1\_female4\_liverec\_250ms\_1m\_sim\_4} \\ \hline
 \raisebox{-0.2ex}[0cm][0cm]{Speech 3}	& \raisebox{-0.2ex}[0cm][0cm]{dev1\_male4\_liverec\_250ms\_1m\_sim\_1\texttt{/}dev1\_male4\_liverec\_250ms\_1m\_sim\_2} \\ \hline
 \raisebox{-0.2ex}[0cm][0cm]{Speech 4}	& \raisebox{-0.2ex}[0cm][0cm]{dev1\_male4\_liverec\_250ms\_1m\_sim\_3\texttt{/}dev1\_male4\_liverec\_250ms\_1m\_sim\_4} \\ \hline
 \raisebox{-0.2ex}[0cm][0cm]{Speech 5}	& \raisebox{-0.2ex}[0cm][0cm]{dev1\_female4\_liverec\_250ms\_1m\_sim\_1\texttt{/}dev1\_male4\_liverec\_250ms\_1m\_sim\_2} \\ \hline
 \raisebox{-0.2ex}[0cm][0cm]{Speech 6}	& \raisebox{-0.2ex}[0cm][0cm]{dev2\_female4\_liverec\_250ms\_1m\_sim\_3\texttt{/}dev1\_male4\_liverec\_250ms\_1m\_sim\_4} \\ \hline
 \raisebox{-0.2ex}[0cm][0cm]{Speech 7}	& \raisebox{-0.2ex}[0cm][0cm]{dev2\_female4\_liverec\_250ms\_1m\_sim\_1\texttt{/}dev1\_female4\_liverec\_250ms\_1m\_sim\_2} \\ \hline
 \raisebox{-0.2ex}[0cm][0cm]{Speech 8}	& \raisebox{-0.2ex}[0cm][0cm]{dev2\_female4\_liverec\_250ms\_1m\_sim\_3\texttt{/}dev1\_female4\_liverec\_250ms\_1m\_sim\_4} \\ \hline
 \raisebox{-0.2ex}[0cm][0cm]{Speech 9}	& \raisebox{-0.2ex}[0cm][0cm]{dev2\_male4\_liverec\_250ms\_1m\_sim\_1\texttt{/}dev1\_male4\_liverec\_250ms\_1m\_sim\_2} \\ \hline
 \raisebox{-0.2ex}[0cm][0cm]{Speech 10}	& \raisebox{-0.2ex}[0cm][0cm]{dev2\_male4\_liverec\_250ms\_1m\_sim\_3\texttt{/}dev1\_male4\_liverec\_250ms\_1m\_sim\_4} \\ \hline
 \raisebox{-0.2ex}[0cm][0cm]{Speech 11}	& \raisebox{-0.2ex}[0cm][0cm]{dev2\_male4\_liverec\_250ms\_1m\_sim\_1\texttt{/}dev1\_female4\_liverec\_250ms\_1m\_sim\_2} \\ \hline
 \raisebox{-0.2ex}[0cm][0cm]{Speech 12}	& \raisebox{-0.2ex}[0cm][0cm]{dev2\_male4\_liverec\_250ms\_1m\_sim\_3\texttt{/}dev1\_female4\_liverec\_250ms\_1m\_sim\_4} \\ \hline
\Hline
\end{tabular}
\end{center}
\end{table}
\endgroup

\subsubsection{{Results and discussion}}

\begin{figure*}[t]
    \begin{center}
        \includegraphics[width=0.96\columnwidth]{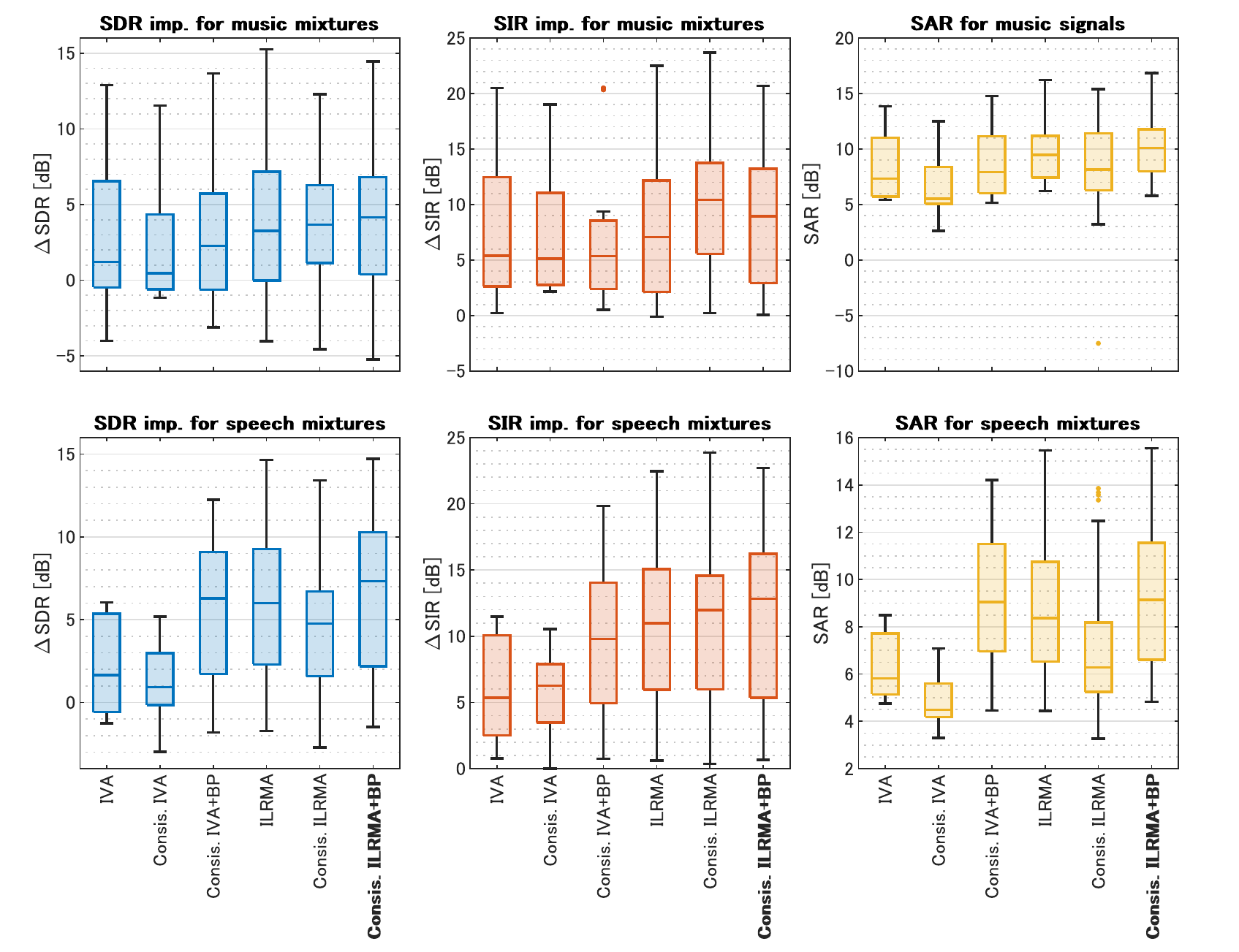}
    \end{center}
    \caption{{SDR improvements (left column), SIR improvements (center column), and SAR (right column) for live-recorded music and speech mixtures, where STFT is performed using the 512-ms-long Hann window with 1/4 shifting. Top row shows the performances for music mixtures and bottom row shows the performances for speech mixtures.}}
    \label{fig:liverecResults}
\end{figure*}

{Figure\:\ref{fig:liverecResults} shows the results of live-recorded music and speech mixtures.
The absolute scores were lower than those for the synthesized mixtures discussed in Sect.~\ref{sec:SynthResults} due to the existence of ambient noise.
Still, we can confirm the improvements of the proposed Consistent IVA+BP and Consistent ILRMA+BP compared to the conventional IVA and ILRMA, respectively, for both the music (upper row) and speech (lower row) mixtures.
In particular, Consistent IVA+BP improved more than 4\:dB over IVA in terms of the median of the $\Delta$SDR of speech mixtures.
Consistent ILRMA+BP achieved the highest performance in terms of the median of the SDR improvement for both music and speech mixtures.
These results confirm that the combination of spectrogram consistency and iterative back projection can assist the separation of determined BSS for a more realistic situation.}

\section{Conclusion}

In this paper, we have proposed a new variant of the state-of-the-art determined BSS algorithm called Consistent ILRMA.
It utilizes the smoothing effect of the inverse STFT in order to assist the separation and enhance the performance.
Experimental results showed that the proposed method can improve the separation performance when the window length is sufficiently large ($\geq$256\:ms in the experimental condition of this paper).
These results demonstrate the potential of considering spectrogram consistency within the state-of-the-art determined BSS algorithm.
{In addition, we experimentally confirmed the importance of iterative back projection for considering spectrogram consistency within determined BSS.}
It should be possible to construct a new source model in consideration of the spectrogram consistency, which can pave the way for the next direction of research on determined BSS.

{\section*{Appendix}}
{\subsection*{Independence between real and imaginary parts of spectrogram}}

\begin{figure}[t]
    \centering
    \subfloat[Hann window]{
    \includegraphics[width=0.93\columnwidth]{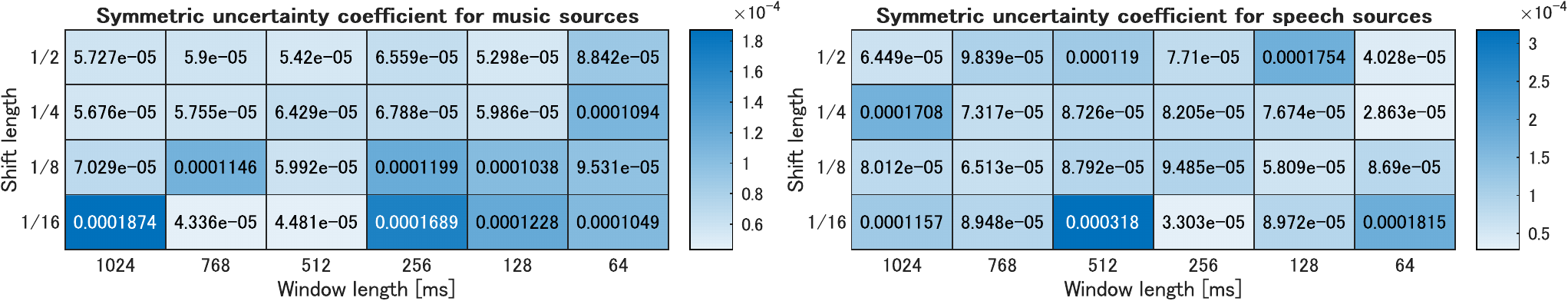}} \\
    \subfloat[Hamming window]{
    \includegraphics[width=0.93\columnwidth]{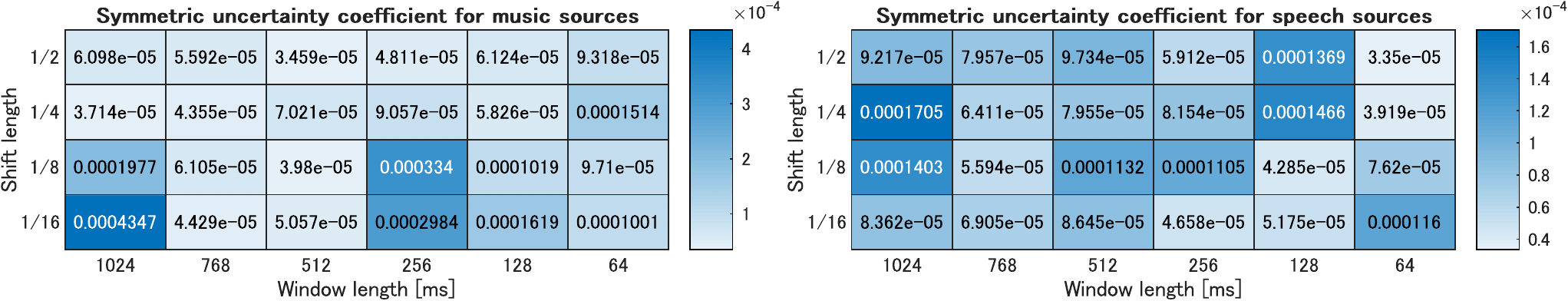}} \\
    \subfloat[Blackman window]{
    \includegraphics[width=0.93\columnwidth]{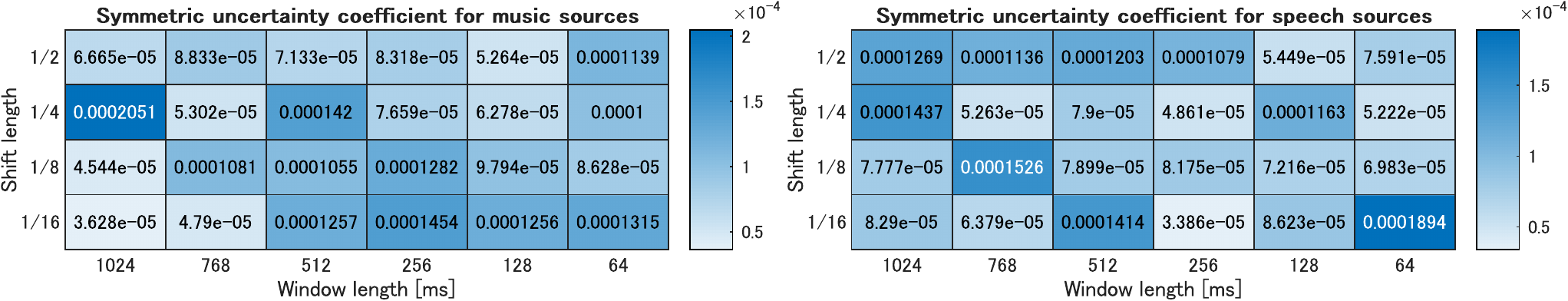}} \\
    \caption{{Symmetric uncertainty coefficient between real and imaginary parts for music and speech sources, where (a) Hann, (b) Hamming, or (c) Blackman window is used in STFT. Left and right columns correspond to the music sources and speech sources, respectively.}}
    \label{fig:symmetUncertCoef}
\end{figure}

{The source generative model \eqref{eq:srcGenModel} assumes that the real and imaginary parts of a source in the time-frequency domain are mutually independent because the generative model has a zero-mean and circularly symmetric shape in the complex plane. 
The independence between real and imaginary parts or amplitude and phase has been investigated, but its validity may depend on the parameters of STFT.
Independence can be measured by a symmetric uncertainty coefficient~\cite{Press1992Book,Andrianakis2009,Mowlaee2020}:
\begin{align}
    C(q_1, q_2) = 2\frac{ H(q_1) + H(q_2) - H(q_1, q_2) }{ H(q_1) + H(q_2) }, \label{eq:symmetUncertCoef}
\end{align}
where $q_1$ and $q_2$ are random variables, $H(q_1)$ and $H(q_2)$ are their entropy, and $H(q_1, q_2)$ is the joint entropy of $q_1$ and $q_2$. 
Since the numerator of \eqref{eq:symmetUncertCoef} corresponds to the mutual information of $q_1$ and $q_2$, the symmetric uncertainty coefficient can be interpreted as normalized mutual information.
When $q_1$ and $q_2$ are mutually independent, \eqref{eq:symmetUncertCoef} becomes zero. 
In contrast, when $q_1$ and $q_2$ are completely dependent, \eqref{eq:symmetUncertCoef} becomes one.

We calculated the symmetric uncertainty coefficient \eqref{eq:symmetUncertCoef} between the real and imaginary parts of a time-frequency bin obtained by applying STFT to music or speech sources. 
Let $s$ be a complex-valued time-frequency bin of a source (the indexes of frequency and time are omitted here). 
The independence between the real and imaginary parts can be measured by $C(\mathrm{Re}(s), \mathrm{Im}(s))$, where $\mathrm{Re}(\cdot)$ and $\mathrm{Im}(\cdot)$ return the real and imaginary parts of an input complex value, respectively.
Here, $H(\mathrm{Re}(s))$, $H(\mathrm{Im}(s))$, and $H(\mathrm{Re}(s), \mathrm{Im}(s))$ were approximately obtained by calculating the histograms of $\mathrm{Re}(s)$ and $\mathrm{Im}(s)$. 
The number of bins in the histograms was set to $10000$.
We used the dry sources listed in Table~\ref{usedSources}: 15 music (instrumental) and eight speech sources. 
The parameters of STFT were the same as those in Sect.~\ref{sec:experiment}.

Figure\:\ref{fig:symmetUncertCoef} shows the symmetric uncertainty coefficients averaged over all bins and sources.
Their values $C(\mathrm{Re}(s), \mathrm{Im}(s))$ were almost zero for all STFT conditions and source types (music or speech), and thus the assumption of independence between real and imaginary parts is valid for music and speech sources.
This fact leads to the generative model assumed in ILRMA.
Note that those symmetric uncertainty coefficients validated the independence of real and imaginary parts at each time-frequency bin.
That is, the inter-bin relation is not considered here.
The proposed method captures such inter-bin relations imposed by the spectrogram consistency, which is not apparent in these bin-wise assessments of independence.}

{\subsection*{Additional experimental results for synthesized mixtures}}

\begin{figure*}[p]
    \begin{center}
        \includegraphics[width=0.96\columnwidth]{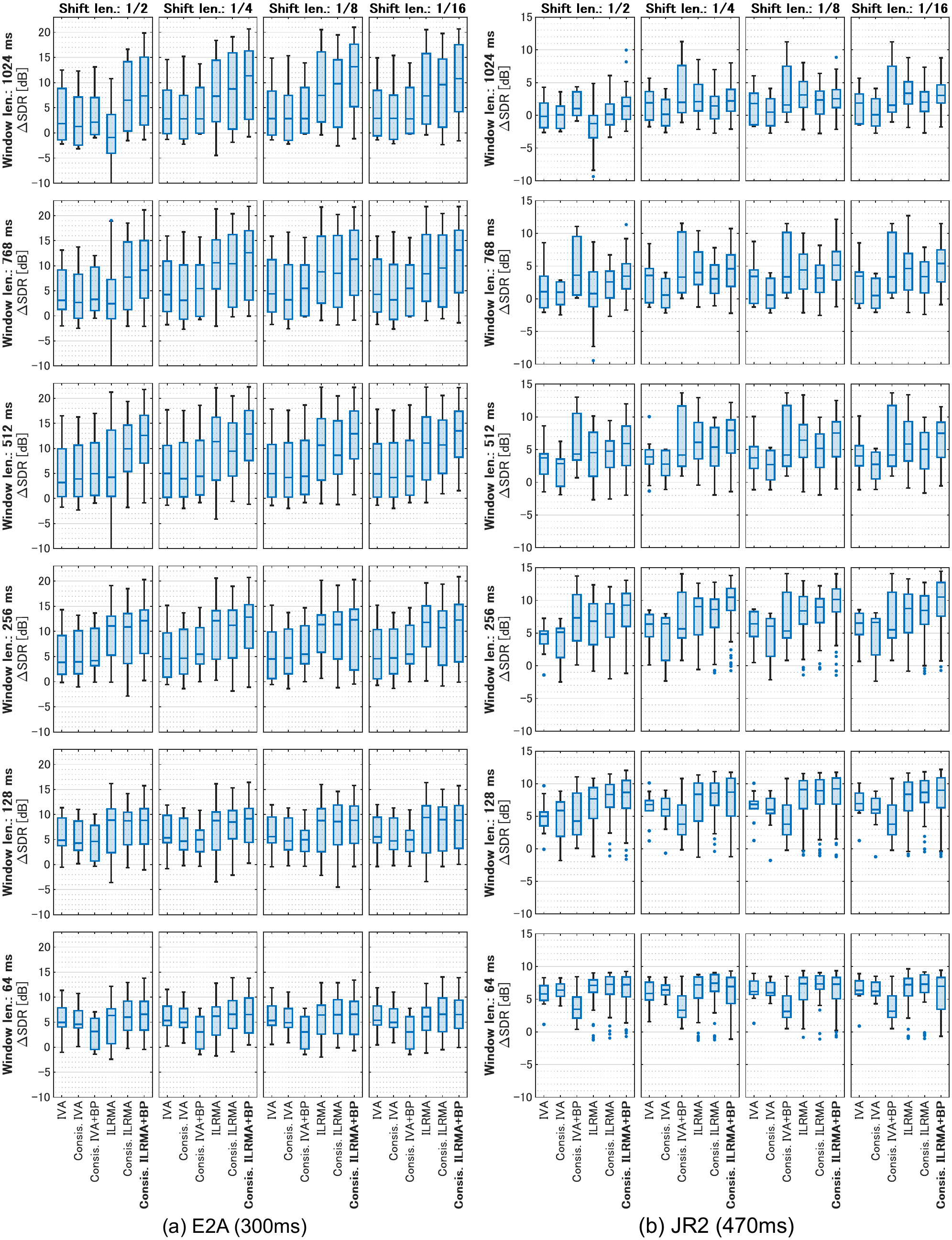}
    \end{center}
    \caption{Average SDR improvements for synthesized music mixtures (Music 1--10) with (a) \texttt{E2A} and (b) \texttt{JR2}, where Hamming window is used in STFT.}
    \label{fig:sdr_Music_hamming}
\end{figure*}

\begin{figure*}[p]
    \begin{center}
        \includegraphics[width=0.96\columnwidth]{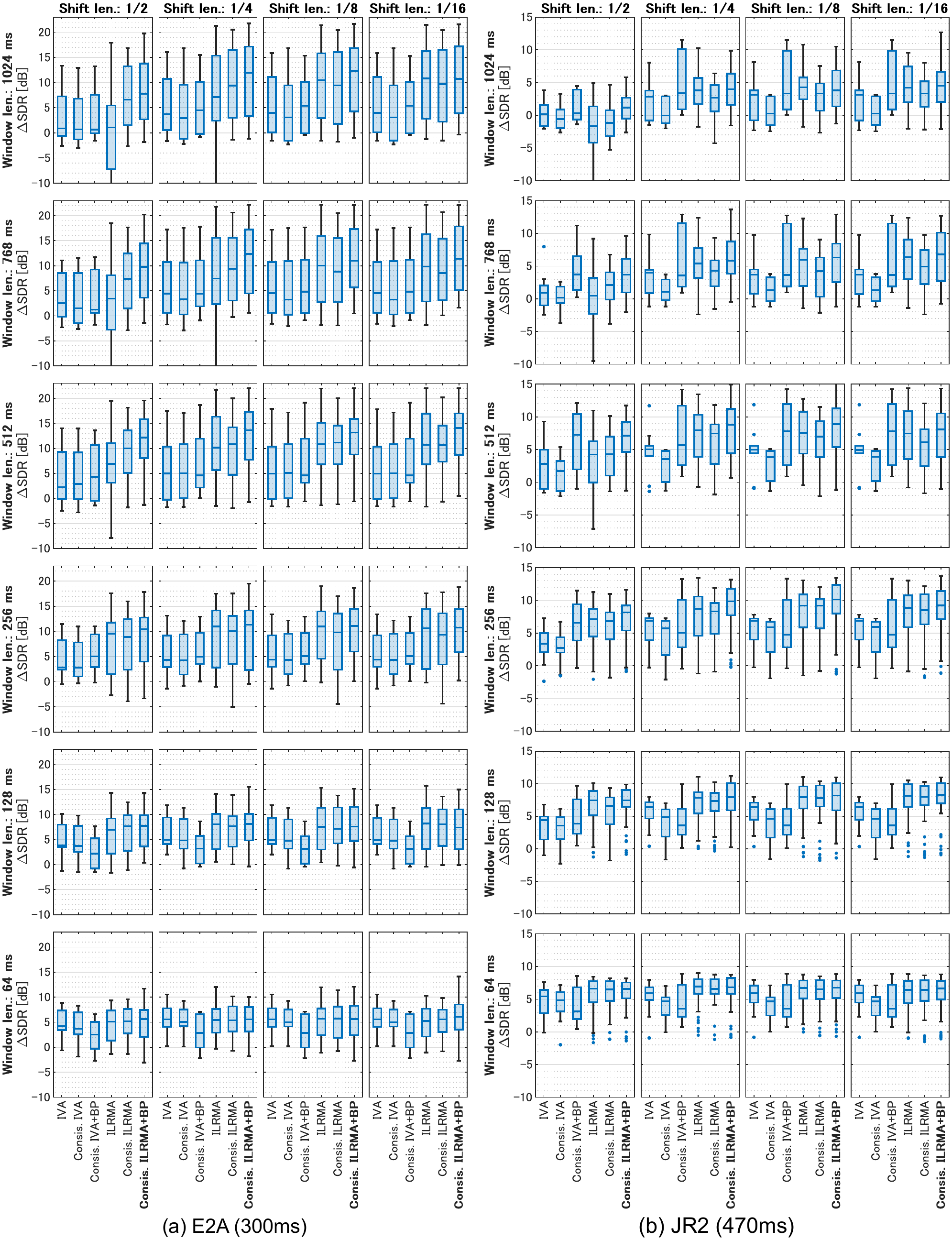}
    \end{center}
    \caption{Average SDR improvements for synthesized music mixtures (Music 1--10) with (a) \texttt{E2A} and (b) \texttt{JR2}, where Blackman window is used in STFT.}
    \label{fig:sdr_Music_blackman}
\end{figure*}

\begin{figure*}[p]
    \begin{center}
        \includegraphics[width=0.96\columnwidth]{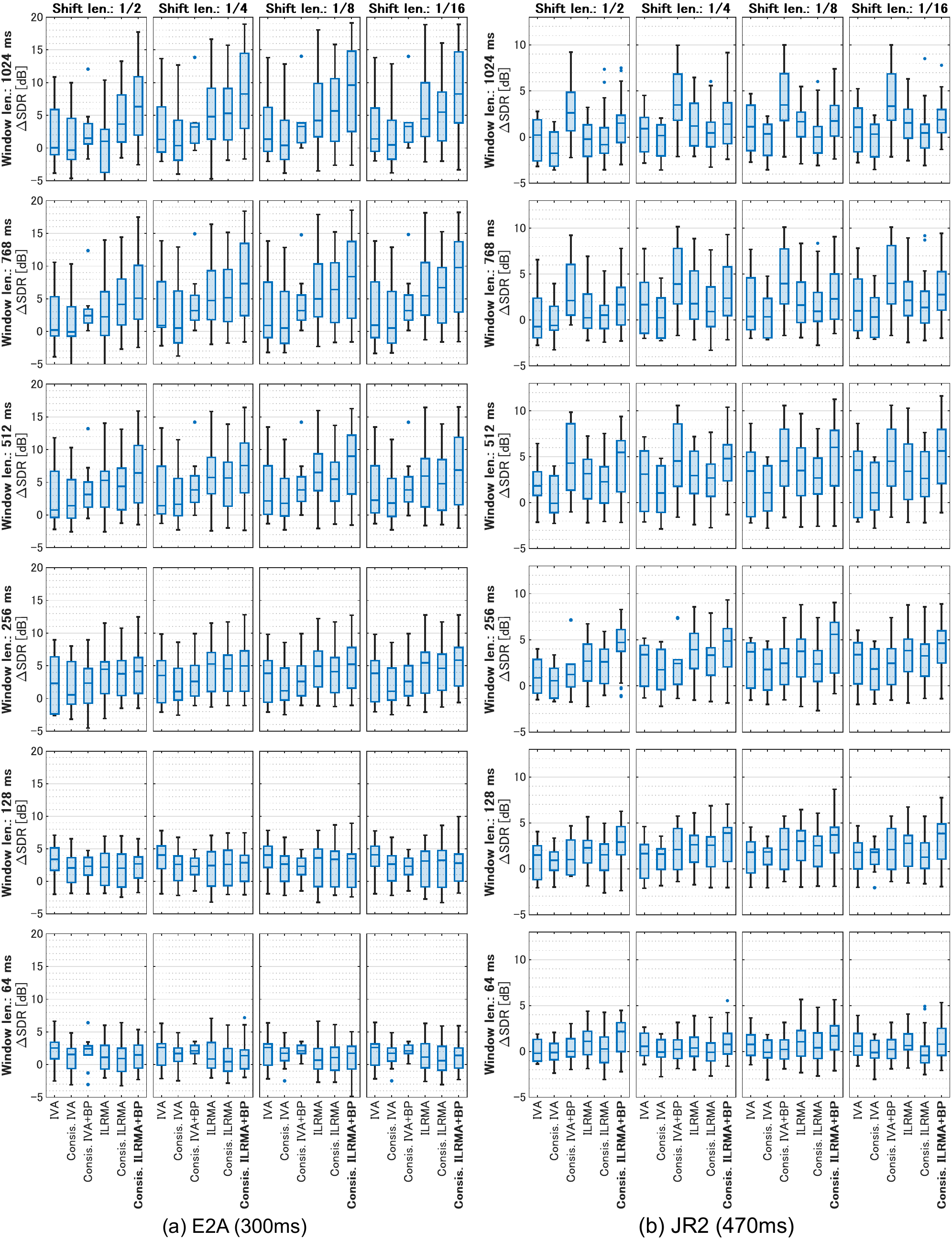}
    \end{center}
    \caption{Average SDR improvements for synthesized speech mixtures (Speech 1--10) with (a) \texttt{E2A} and (b) \texttt{JR2}, where Hamming window is used in STFT.}
    \label{fig:sdr_Speech_hamming}
\end{figure*}

\begin{figure*}[p]
    \begin{center}
        \includegraphics[width=0.96\columnwidth]{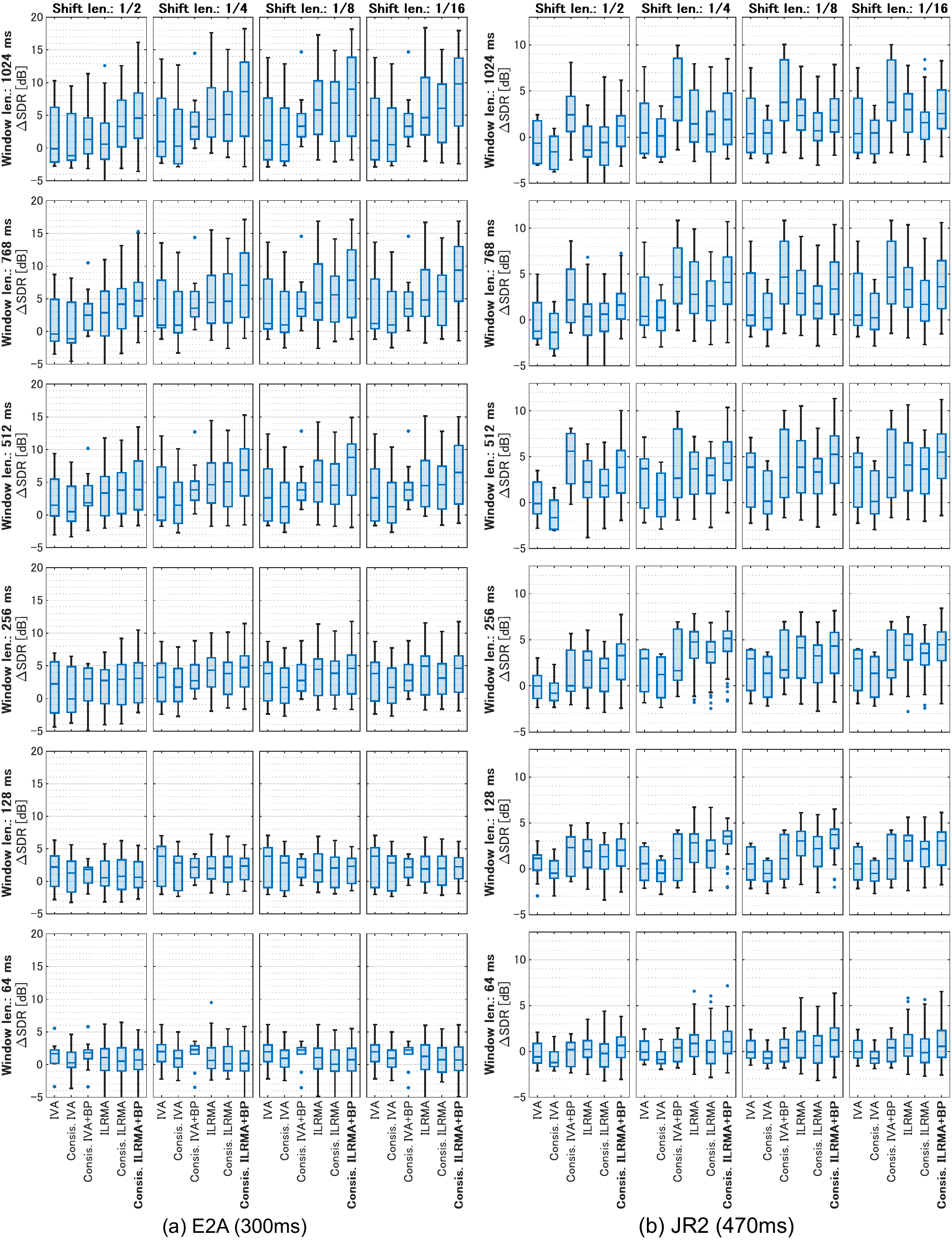}
    \end{center}
    \caption{Average SDR improvements for synthesized speech mixtures (Speech 1--10) with (a) \texttt{E2A} and (b) \texttt{JR2}, where Blackman window is used in STFT.}
    \label{fig:sdr_Speech_blackman}
\end{figure*}

\begin{figure*}[p]
    \begin{center}
        \includegraphics[width=0.96\columnwidth]{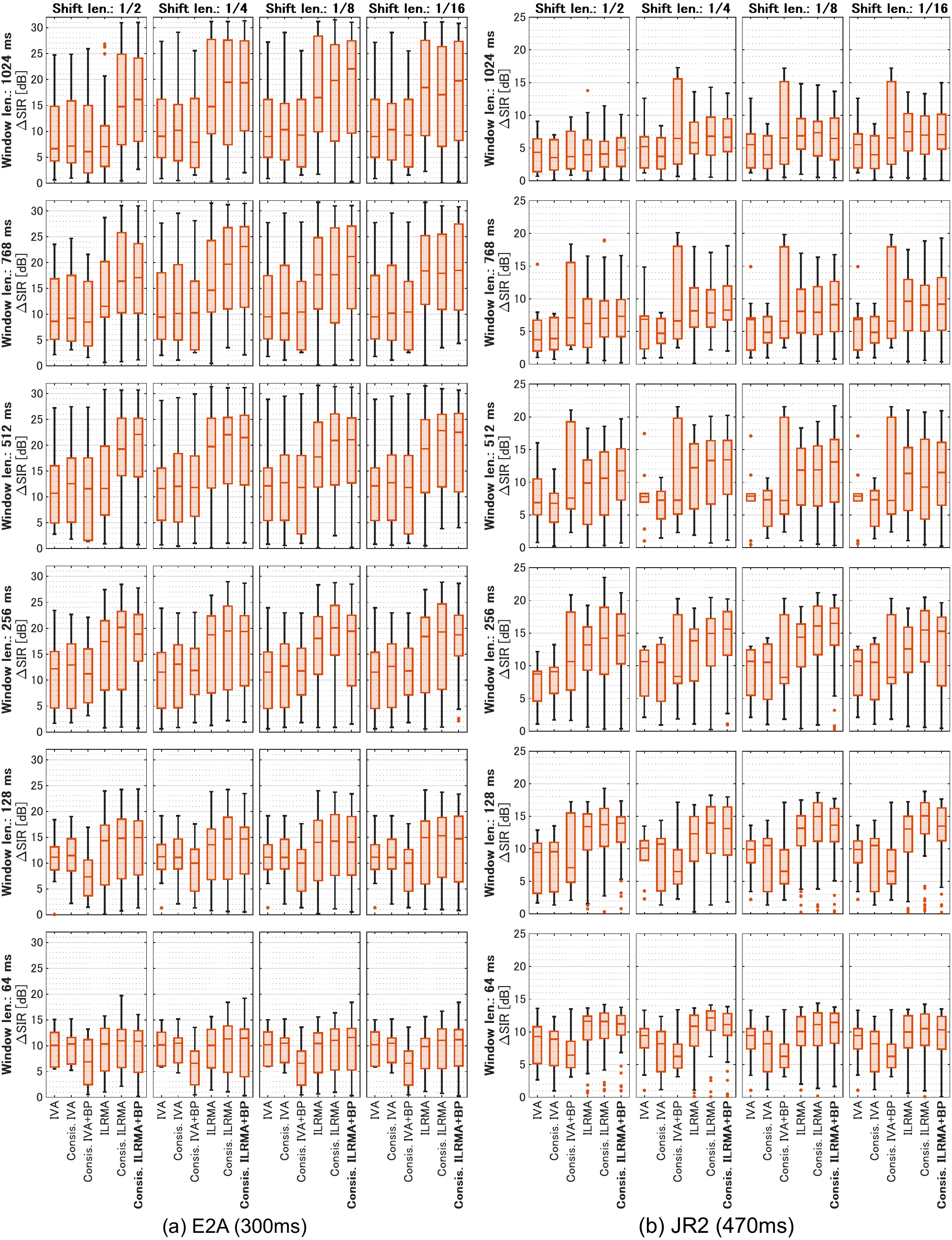}
    \end{center}
    \caption{Average SIR improvements for synthesized music mixtures (Music 1--10) with (a) \texttt{E2A} and (b) \texttt{JR2}, where Hann window is used in STFT.}
    \label{fig:sir_Music_hann}
\end{figure*}

\begin{figure*}[p]
    \begin{center}
        \includegraphics[width=0.96\columnwidth]{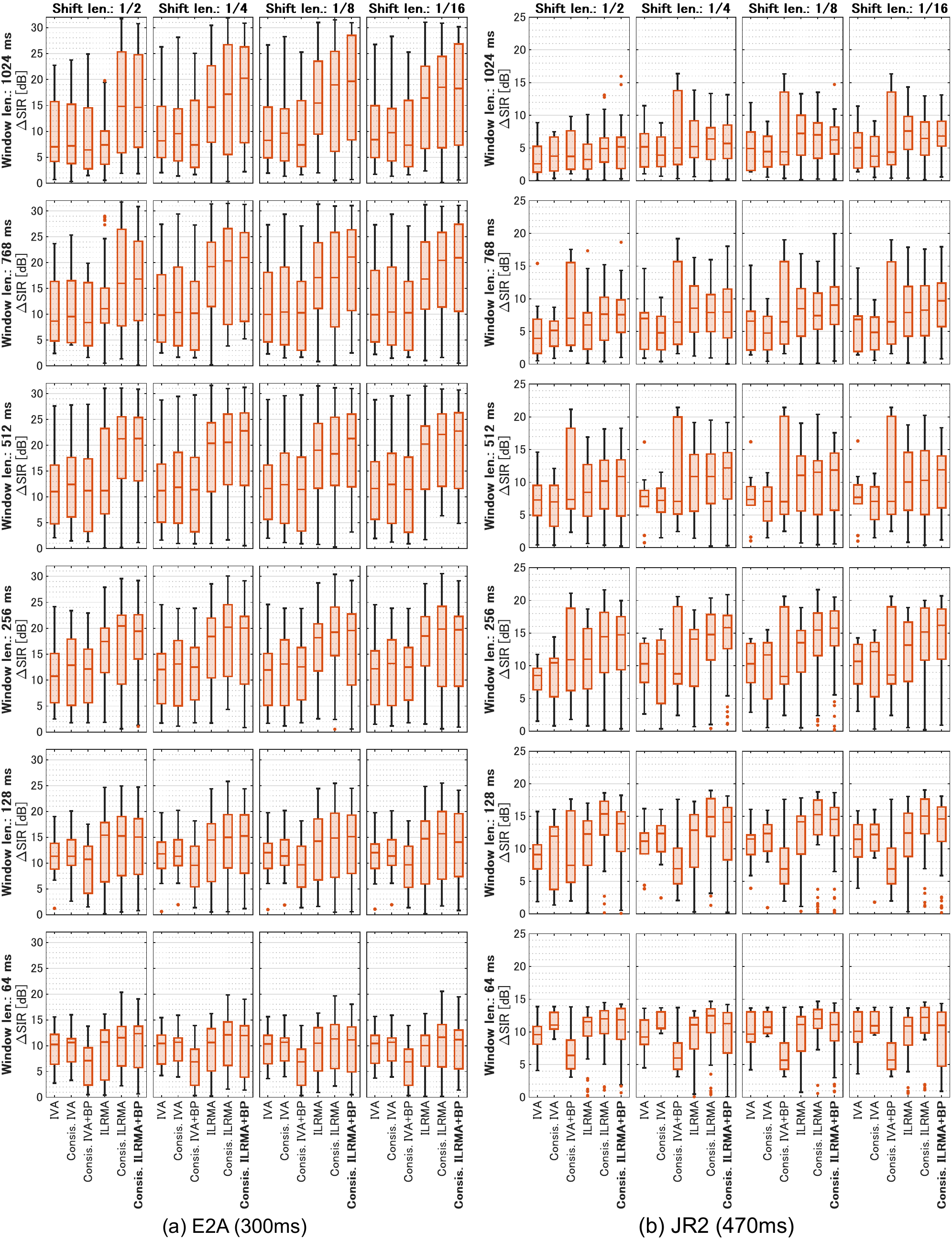}
    \end{center}
    \caption{Average SIR improvements for synthesized music mixtures (Music 1--10) with (a) \texttt{E2A} and (b) \texttt{JR2}, where Hamming window is used in STFT.}
    \label{fig:sir_Music_hamming}
\end{figure*}

\begin{figure*}[p]
    \begin{center}
        \includegraphics[width=0.96\columnwidth]{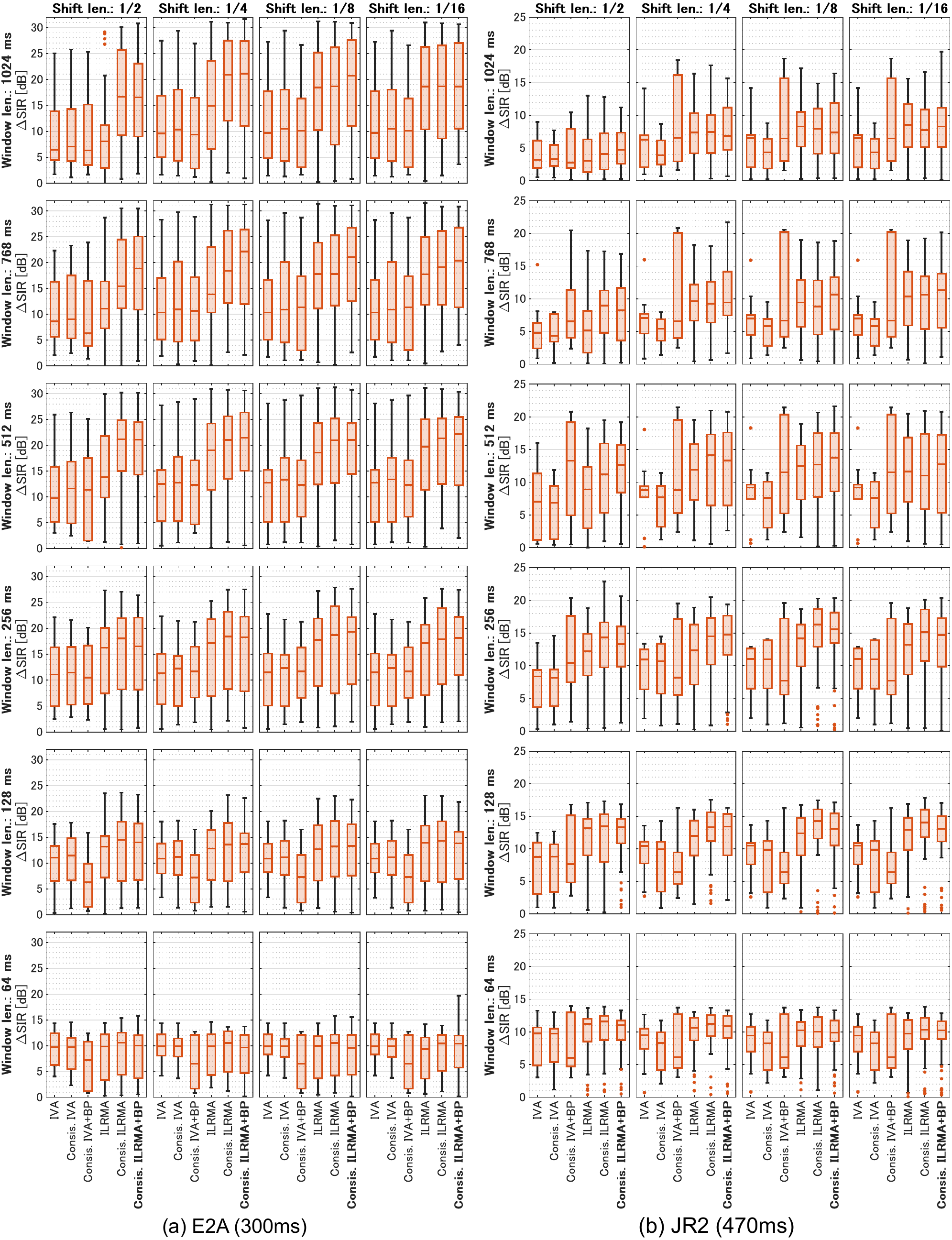}
    \end{center}
    \caption{Average SIR improvements for synthesized music mixtures (Music 1--10) with (a) \texttt{E2A} and (b) \texttt{JR2}, where Blackman window is used in STFT.}
    \label{fig:sir_Music_blackman}
\end{figure*}

\begin{figure*}[p]
    \begin{center}
        \includegraphics[width=0.96\columnwidth]{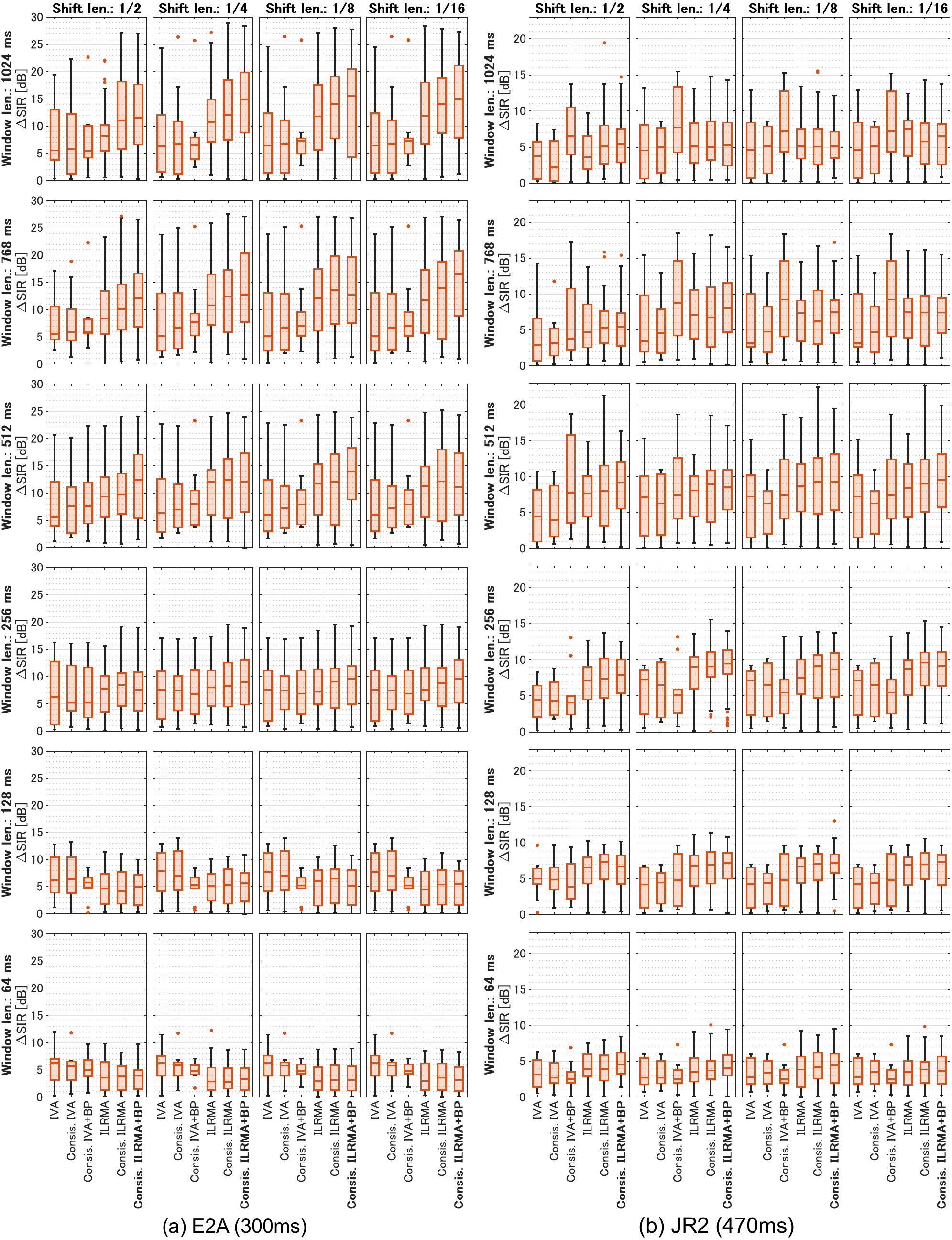}
    \end{center}
    \caption{Average SIR improvements for synthesized speech mixtures (Speech 1--10) with (a) \texttt{E2A} and (b) \texttt{JR2}, where Hann window is used in STFT.}
    \label{fig:sir_Speech_hann}
\end{figure*}

\begin{figure*}[p]
    \begin{center}
        \includegraphics[width=0.96\columnwidth]{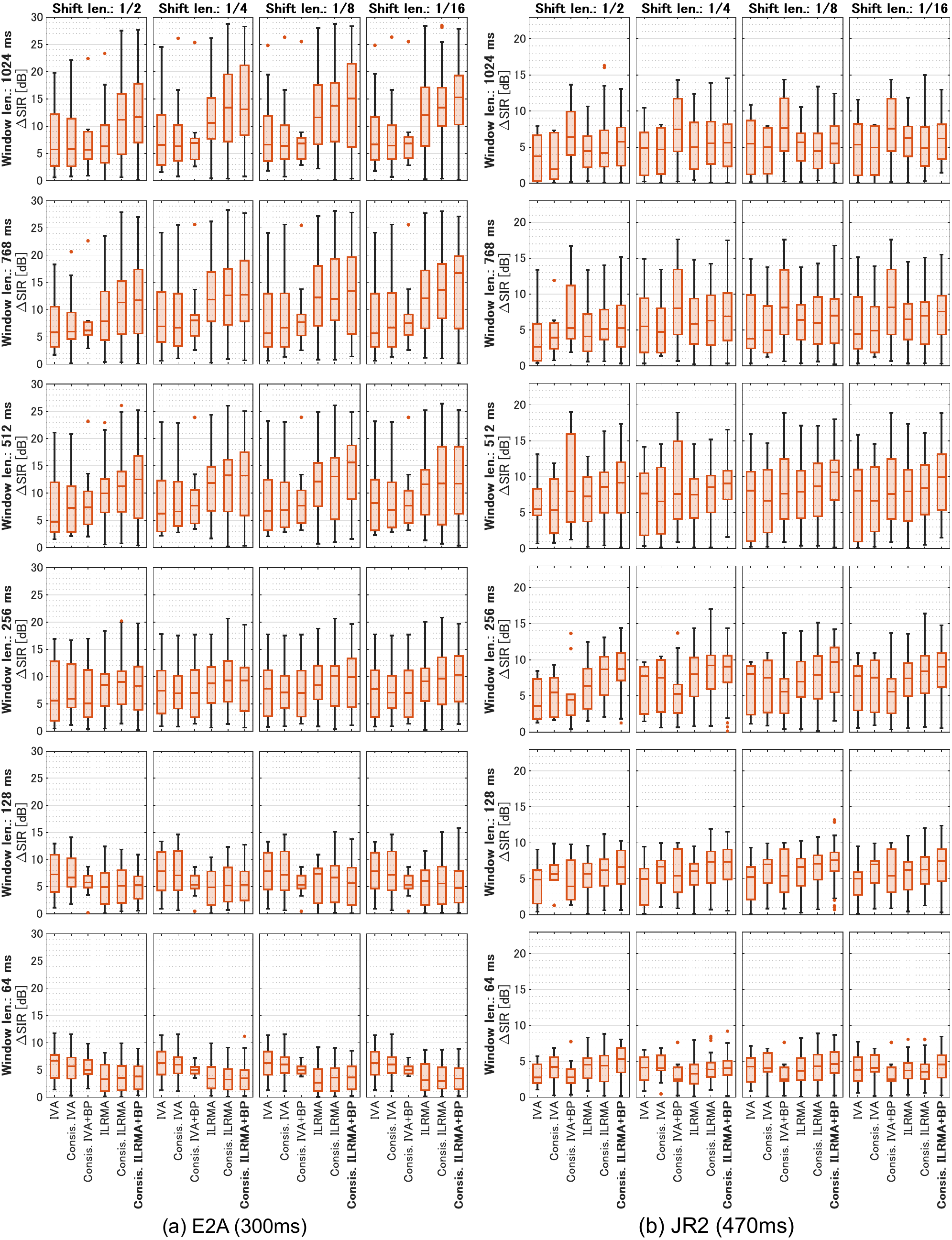}
    \end{center}
    \caption{Average SIR improvements for synthesized speech mixtures (Speech 1--10) with (a) \texttt{E2A} and (b) \texttt{JR2}, where Hamming window is used in STFT.}
    \label{fig:sir_Speech_hamming}
\end{figure*}

\begin{figure*}[p]
    \begin{center}
        \includegraphics[width=0.96\columnwidth]{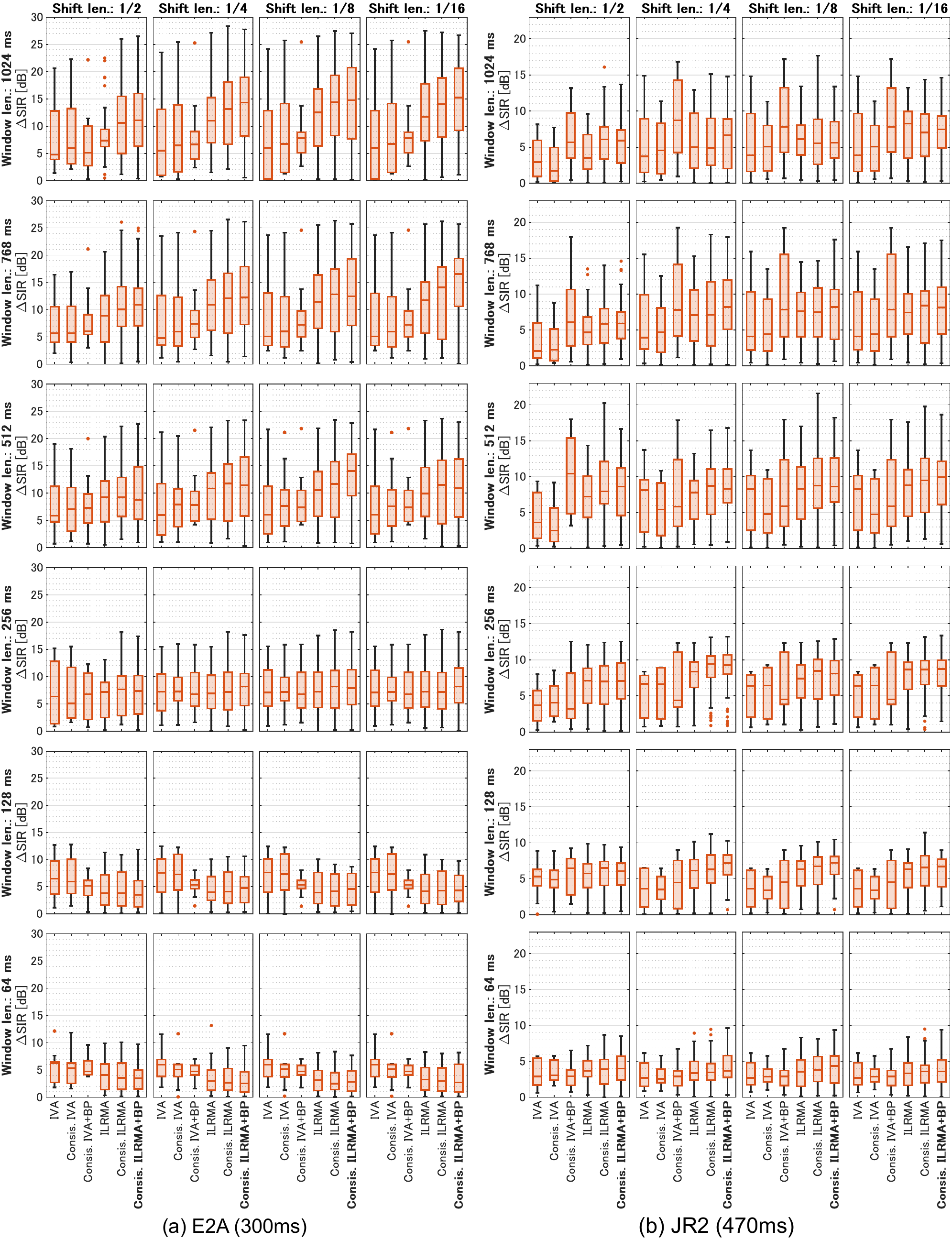}
    \end{center}
    \caption{Average SIR improvements for synthesized speech mixtures (Speech 1--10) with (a) \texttt{E2A} and (b) \texttt{JR2}, where Blackman window is used in STFT.}
    \label{fig:sir_Speech_blackman}
\end{figure*}

\begin{figure*}[p]
    \begin{center}
        \includegraphics[width=0.96\columnwidth]{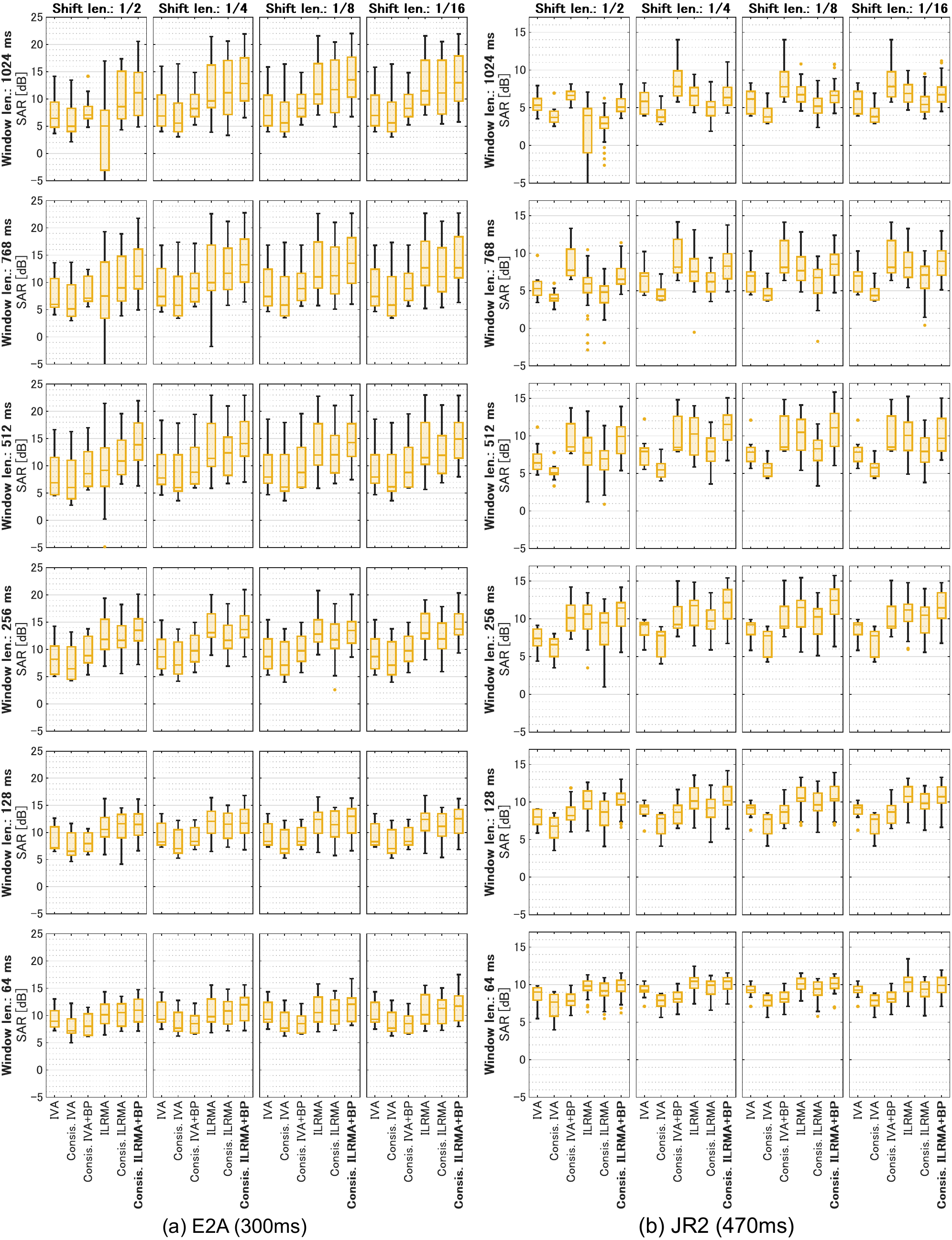}
    \end{center}
    \caption{Average SAR for synthesized music mixtures (Music 1--10) with (a) \texttt{E2A} and (b) \texttt{JR2}, where Hann window is used in STFT.}
    \label{fig:sar_Music_hann}
\end{figure*}

\begin{figure*}[p]
    \begin{center}
        \includegraphics[width=0.96\columnwidth]{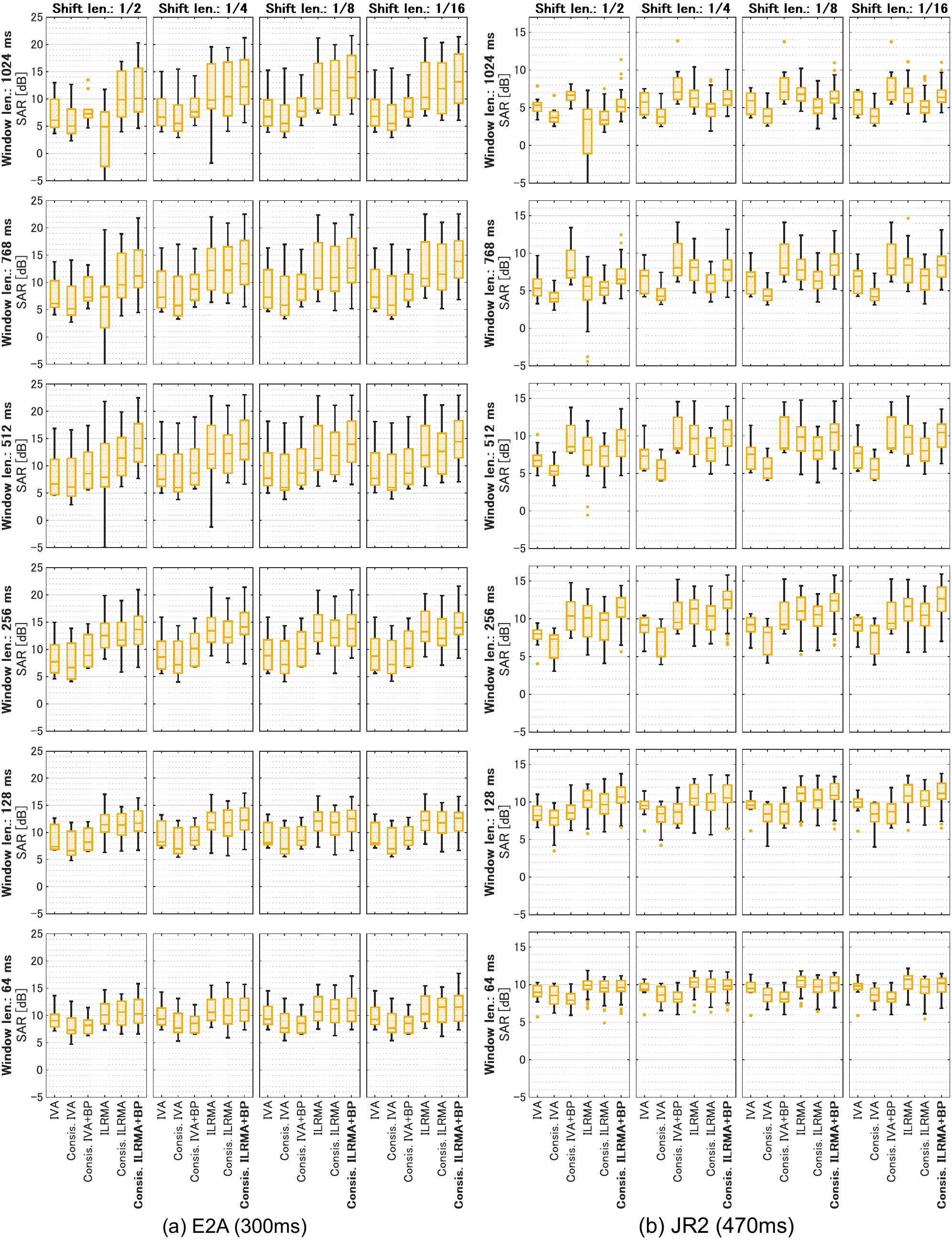}
    \end{center}
    \caption{Average SAR for synthesized music mixtures (Music 1--10) with (a) \texttt{E2A} and (b) \texttt{JR2}, where Hamming window is used in STFT.}
    \label{fig:sar_Music_hamming}
\end{figure*}

\begin{figure*}[p]
    \begin{center}
        \includegraphics[width=0.96\columnwidth]{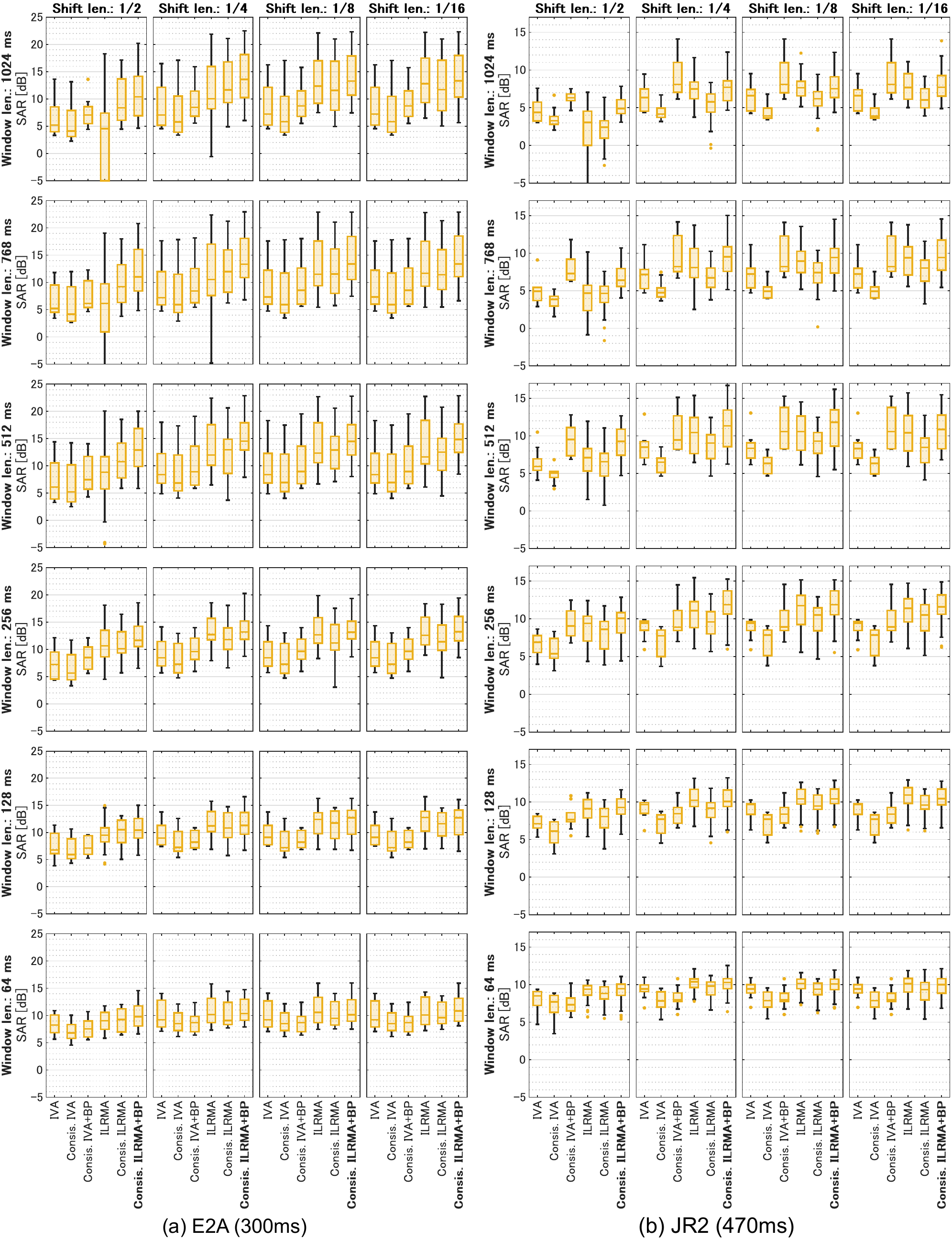}
    \end{center}
    \caption{Average SAR for synthesized music mixtures (Music 1--10) with (a) \texttt{E2A} and (b) \texttt{JR2}, where Blackman window is used in STFT.}
    \label{fig:sar_Music_blackman}
\end{figure*}

\begin{figure*}[p]
    \begin{center}
        \includegraphics[width=0.96\columnwidth]{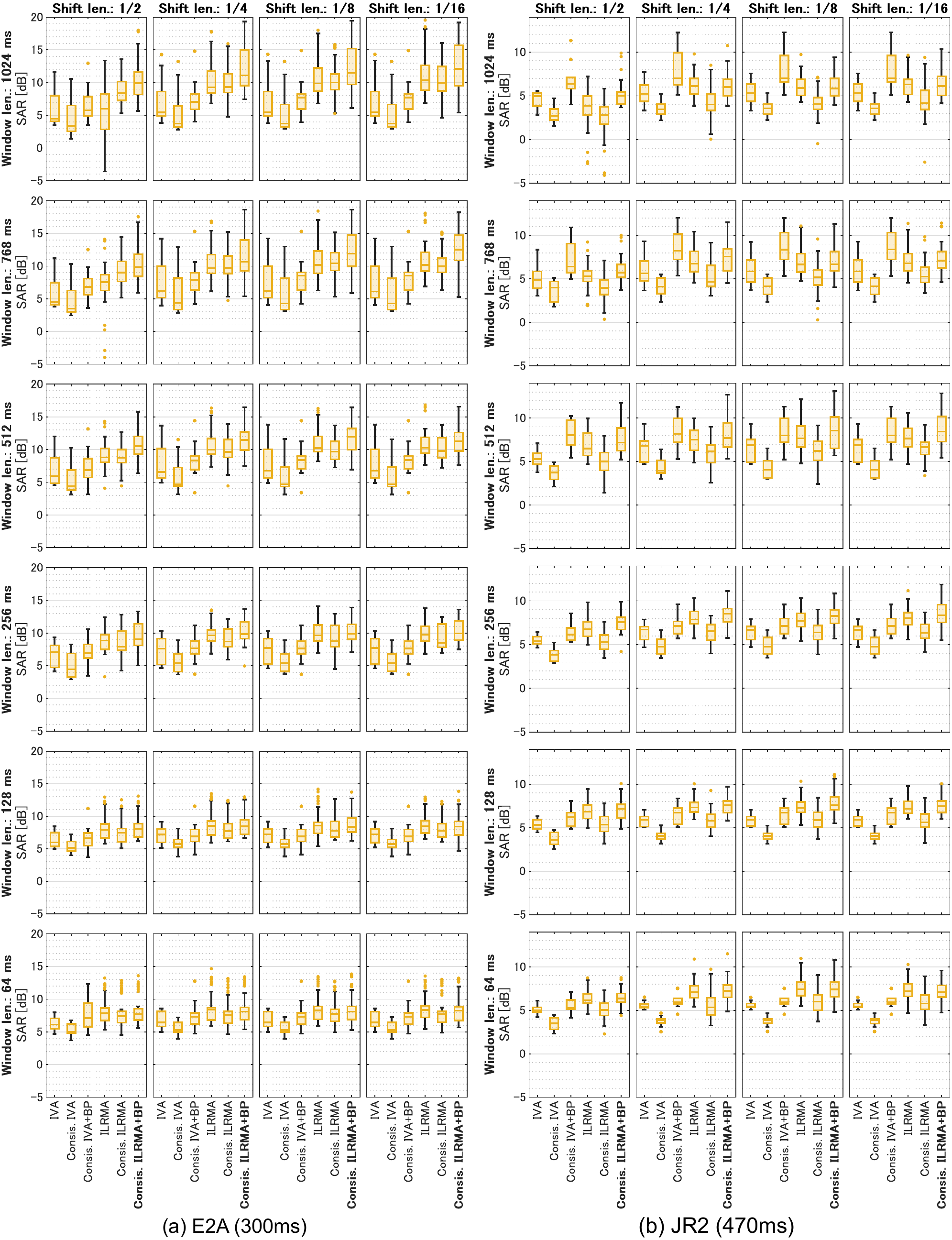}
    \end{center}
    \caption{Average SAR for synthesized speech mixtures (Speech 1--10) with (a) \texttt{E2A} and (b) \texttt{JR2}, where Hann window is used in STFT.}
    \label{fig:sar_Speech_hann}
\end{figure*}

\begin{figure*}[p]
    \begin{center}
        \includegraphics[width=0.96\columnwidth]{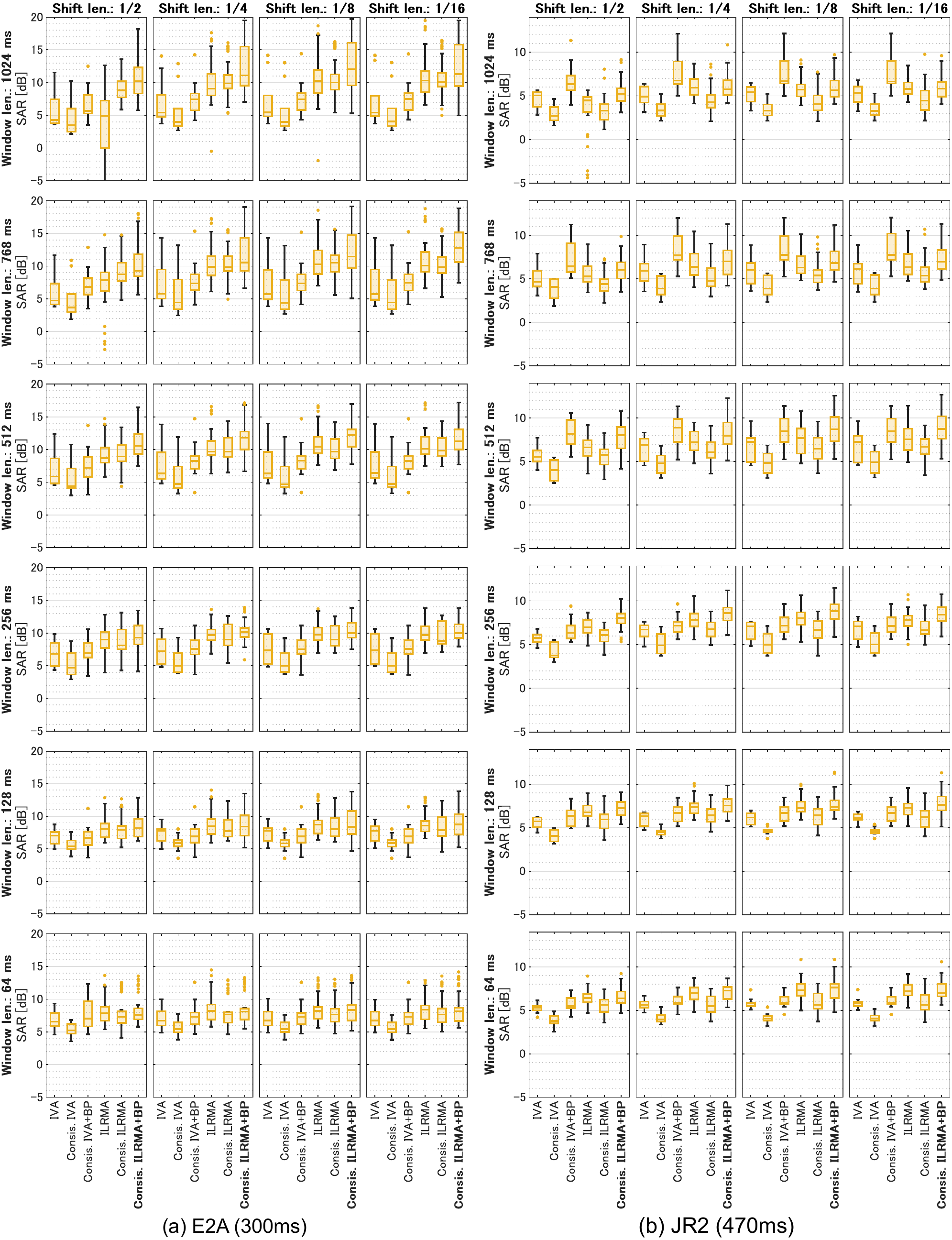}
    \end{center}
    \caption{Average SAR for synthesized speech mixtures (Speech 1--10) with (a) \texttt{E2A} and (b) \texttt{JR2}, where Hamming window is used in STFT.}
    \label{fig:sar_Speech_hamming}
\end{figure*}

\begin{figure*}[p]
    \begin{center}
        \includegraphics[width=0.96\columnwidth]{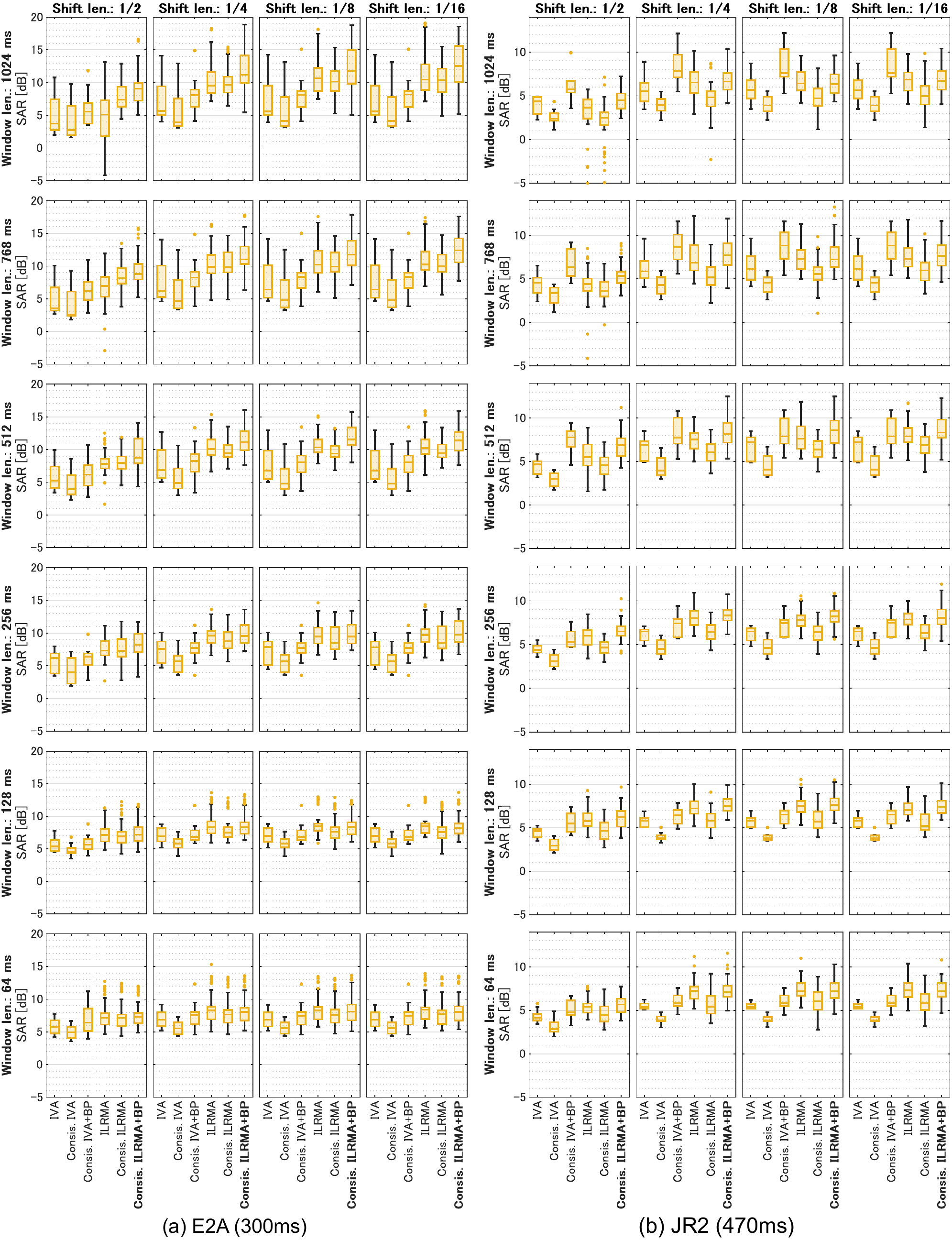}
    \end{center}
    \caption{Average SAR for synthesized speech mixtures (Speech 1--10) with (a) \texttt{E2A} and (b) \texttt{JR2}, where Blackman window is used in STFT.}
    \label{fig:sar_Speech_blackman}
\end{figure*}

{Figures\:\ref{fig:sdr_Music_hamming}--\ref{fig:sdr_Speech_blackman}, \ref{fig:sir_Music_hann}--\ref{fig:sir_Speech_blackman}, and \ref{fig:sar_Music_hann}--\ref{fig:sar_Speech_blackman} show the SDR improvements, SIR improvements, and SAR, respectively, for synthesized music and speech mixtures. 
These figures correspond to the results and discussions in Sect.~\ref{sec:SynthResults}.}

\begin{backmatter}

\section*{Abbreviations}
  BSS: Blind source separation; ICA: independent component analysis; STFT: short-time Fourier transform; FDICA: frequency-domain independent component analysis; IVA: independent vector analysis; ILRMA: independent low-rank matrix analysis; NMF: nonnegative matrix factorization; IS-NMF: nonnegative matrix factorization based on the Itakura--Saito divergence; SDR: source-to-distortion ratio; SIR: source-to-interference ratio; SAR: sources-to-artificial ratio

\section*{Availability of data and material}
  The datasets used for the experiments in this paper are openly available: SiSEC 2011 (\url{http://sisec2011.wiki.irisa.fr/}) and RWCP-SSD (\url{http://research.nii.ac.jp/src/en/RWCP-SSD.html}).
  Our MATLAB implementation of the proposed method is also openly available at the following site: \url{https://github.com/d-kitamura/ILRMA/blob/master/consistentILRMA.m}

\section*{Competing interests}
  The authors declare that they have no competing interests.

\section*{Funding}
  This work was partially supported by JSPS Grants-in-Aid for Scientific Research 19K20306 and 19H01116.

\section*{Authors' contributions}
  DK derived the algorithm, performed the experiment, drafted the manuscript for initial submission, and revised the manuscript. KY proposed the main idea, gave advice, mainly wrote the manuscript for initial submission, and corrected the draft of revised manuscript.

\section*{Acknowledgements}
  The authors would like to thank Nao Toshima for his support on the experiment.
  Also, the authors would like to thank the anonymous reviewers for their valuable comments and suggestions that helped improve the quality of this manuscript.
  

\bibliographystyle{bmc-mathphys} 

\end{backmatter}
\end{document}